\documentclass[]{aastex631}
\usepackage{amsmath}

\newcommand{\msun}{\textrm{M}_\odot}

\newcommand\pupa{RX J0822.0$-$4300}
\newcommand\velajr{CXOU J085201.4$-$461753}
\newcommand\twelveohseven{1E 1207.4$-$5209}
\newcommand\gthreethirty{CXOU J160103.1$-$513353}
\newcommand\gthreefortyseven{1WGA J1713.4$-$3949}
\newcommand\gthreefifty{XMMU J172054.5$-$372652}
\newcommand\kesseventynine{CXOU J185238.6+004020}
\newcommand\casa{CXOU J232327.9+584842}

\newcommand\gthreefiftythree{XMMU J173203.3$-$344518}
\newcommand\gfifteen{CXOU J181852.0$-$150213}

\newcommand\velajrsnr{G266.1$-$1.2 (Vela Jr.)}
\newcommand\gthreefortysevensnr{G347.3$-$0.5}
\newcommand\kesseventyninesnr{Kes 79}
\newcommand\pupasnr{Puppis~A}
\newcommand\twelveohsevensnr{PKS 1209$-$51/52}
\newcommand\gthreethirtysnr{G330.2+1.0}
\newcommand\gthreefiftysnr{G350.1$-$0.3}
\newcommand\casasnr{Cas~A}
\newcommand\gthreefiftythreesnr{G353.6$-$0.7}
\newcommand\gfifteensnr{G15.9$+$0.2}

\newcommand\xmm{XMM-Newton}
\newcommand\chandra{Chandra}

\shorttitle{Do CCOs have Carbon Atmospheres?}
\shortauthors{Alford and Halpern}

\begin{document}
\title{Do Central Compact Objects have Carbon Atmospheres?}

\correspondingauthor{Jason A. J. Alford}
\email{jason@astro.columbia.edu}

\author[0000-0002-2312-8539]{J. A. J. Alford}
\affil{Columbia Astrophysics Laboratory, Columbia University, 550 West 120th Street, New York NY, 10027, USA}
\author[0000-0003-4814-2377]{J. P. Halpern}
\affil{Columbia Astrophysics Laboratory, Columbia University, 550 West 120th Street, New York NY, 10027, USA}

\begin{abstract}
Only three of the dozen central compact objects (CCOs) in supernova remnants (SNRs) show thermal X-ray pulsations due to non-uniform surface temperature (hot-spots). 
The absence of X-ray pulsations from several unpulsed CCOs has motivated suggestions that they have uniform-temperature carbon atmospheres (UTCAs), which adequately fit their spectra with  appropriate neutron star (NS) surface areas.
This is in contrast to the two-temperature blackbody or hydrogen atmospheres that also fit well.  
Here we investigate the applicability of UTCAs to CCOs.
We show the following: 
(i) The phase-averaged spectra of the three pulsed CCOs can also be fitted with a UTCA of the appropriate NS area, despite pulsed CCOs manifestly having non-uniform surface temperature. 
A good spectral fit is therefore not strong support for the UTCA model of unpulsed CCOs.  
(ii) An improved spectrum of one unpulsed CCO, previously analyzed with a UTCA, does not allow an acceptable fit.  
(iii) For two unpulsed CCOs, the UTCA does not allow a distance compatible with the SNR distance.
These results imply that, in general, CCOs must have hot, localized regions on the NS surface.
We derive new X-ray pulse modulation upper limits on the unpulsed CCOs, and constrain their hot spot sizes and locations.  
We develop an alternative model that accounts for both the pulsed and unpulsed CCOs: a range of angles between hot spot and rotation axes consistent with an exponential distribution with scale factor $\lambda \sim 20^{\circ}$.
We discuss physical mechanisms that could produce such small angles and small hot-spots.
\end{abstract}

\section{Introduction}

\subsection{Central Compact Objects}
Central Compact Objects (CCOs) are a class of young, isolated  neutron stars (NSs) found in supernova remnants (SNRs).
Their defining characteristics are steady, thermal X-ray emission, non-detection at all other wavelengths, and the absence of a surrounding pulsar wind nebula. 
The similar number of CCOs in SNRs, relative to other classes of young NSs, suggests that CCOs birth rates are a significant fraction of all NS births.
See \cite{del08} and \cite{del17} for reviews of CCOs.

X-ray pulsations have been detected from three of the eight confirmed CCOs, located in the \pupasnr, \kesseventyninesnr,  and \twelveohsevensnr \ SNRs \citep{zav00,got05,got09}.
\cite{hal10a} and \cite{got13} measured the period derivatives $\dot{P}$ of these three CCO pulsars. 
The implied dipole surface magnetic field strengths are $B_s = (2.9,3.1,9.8)\times10^{10}$~G, for the CCOs in \pupasnr, \kesseventyninesnr, and \twelveohsevensnr, respectively.
The almost identical spin-down measured magnetic fields for the CCOs in \kesseventyninesnr \ and \pupasnr \ are the smallest ever measured in young NSs.

Their X-ray pulsations indicate that at least these three CCOs have localized regions on their surfaces producing their observed thermal radiation. 
Spectral fits and modeling also indicate the need for more than one surface temperature \citep{got09,hal10a,big03,got10,bog14,alf22}.
The X-ray flux from the hot-spots exceeds their spin-down power, and is likely supplied by residual cooling \citep{got13}.
Strong crustal magnetic fields, possibly toroidal or quadrupolar components, are required to conduct heat toward these small regions on the NS surface \citep{gre83}.

The CCOs without detected pulsations, having similar spectral properties as the CCO pulsars, are also likely NSs born with weak ($10^{10-11}$~G) dipole magnetic fields. They may also have hot, localized surface regions, although producing lower amplitude pulsations that cannot be detected in existing data.
This could be due to a combination of unfavorable viewing angles, and hot-spots that are located close to their rotational poles (see e.g. \citealt{bog14,alf22}).
Alternatively, these CCOs could have uniform-temperature surfaces, which would not modulate the X-ray flux as the NS rotates.  
But this alternative scenario would require a NS atmosphere composed of a mid-Z element such as carbon.
This is because the NS radii implied by fits to one or two-temperature blackbody or hydrogen atmospheres are too small for a NS, while a  single-temperature non-magnetic carbon atmosphere gives a reasonable radius for the CCO in Cas~A \citep{ho09}. 
Following this pioneering suggestion for Cas~A, several more CCOs were proposed to have UTCAs. These works will be reviewed in Section~1.4.

\subsection{Carbon from Diffusive Nuclear Burning}
The chemical composition of a NS atmosphere can change over time through accretion of circumstellar material, spallation, and diffusive nuclear burning (DNB).
Accretion and spallation increase the fraction of hydrogen and helium in the atmosphere.
In contrast, DNB decreases the hydrogen and helium fraction via nuclear fusion into heavier elements such as carbon and oxygen.
Gravitational settling occurs on the timescale of seconds, so NS photospheres will consist of the lightest elements available \citep{alc80}.

Young NSs have $\sim 10^{6}$~K photospheres, not hot enough to efficiently burn H and He into heavier elements during a NS lifetime.
However, H and He can diffuse down from the atmosphere into the outer neutron star envelope, where the temperature can increase by 1--2 orders of magnitude at a depth of just 10~m below the surface.
The higher temperatures and densities in this region can efficiently burn H and He into heavier elements.
Then, in the absence of further accretion or spallation, the H and He can be depleted from NS atmosphere.
\cite{cha03} and \cite{cha042} developed the theory of DNB, and calculated the H burning timescales.
They found that H on the NS surface could be depleted on timescales as short as $10^{2}-10^{4}$ yr.
\cite{cha10} extended the theory of diffusive nuclear burning to include He burning, and found that, in the absence of further accretion, He could also be efficiently depleted from the atmospheres of young, hot NSs. 
This efficient H and He burning could then produce a NS with a carbon atmosphere.
Since DNB is more efficient at higher temperatures, it is expected to be most effective in younger NSs, such as the Cas A CCO, which have not had much time to cool.

The mass of atmospheric material needed to change the spectrum of a NS is very small, $4\pi R_{\rm NS}^2/\kappa \sim 10^{-20}\ \msun$ assuming $\kappa\sim 1$~cm$^{2}$~g$^{-1}$. But it is impossible to measure the small corresponding accretion rate \citep{cha10} of $\sim10^{-22}\ \msun\ {\rm yr}^{-1}$ directly. 
While the rate of nuclear burning in a NS with a given composition and temperature is well understood, the rate of accretion of hydrogen onto any one particular NS inside a SNR is highly uncertain.
The chemical composition of CCO atmospheres must be determined through direct observation of their X-ray spectra.

\subsection{NS Atmosphere Models}
\begin{figure*}
\centering
\includegraphics[width=0.9\linewidth]{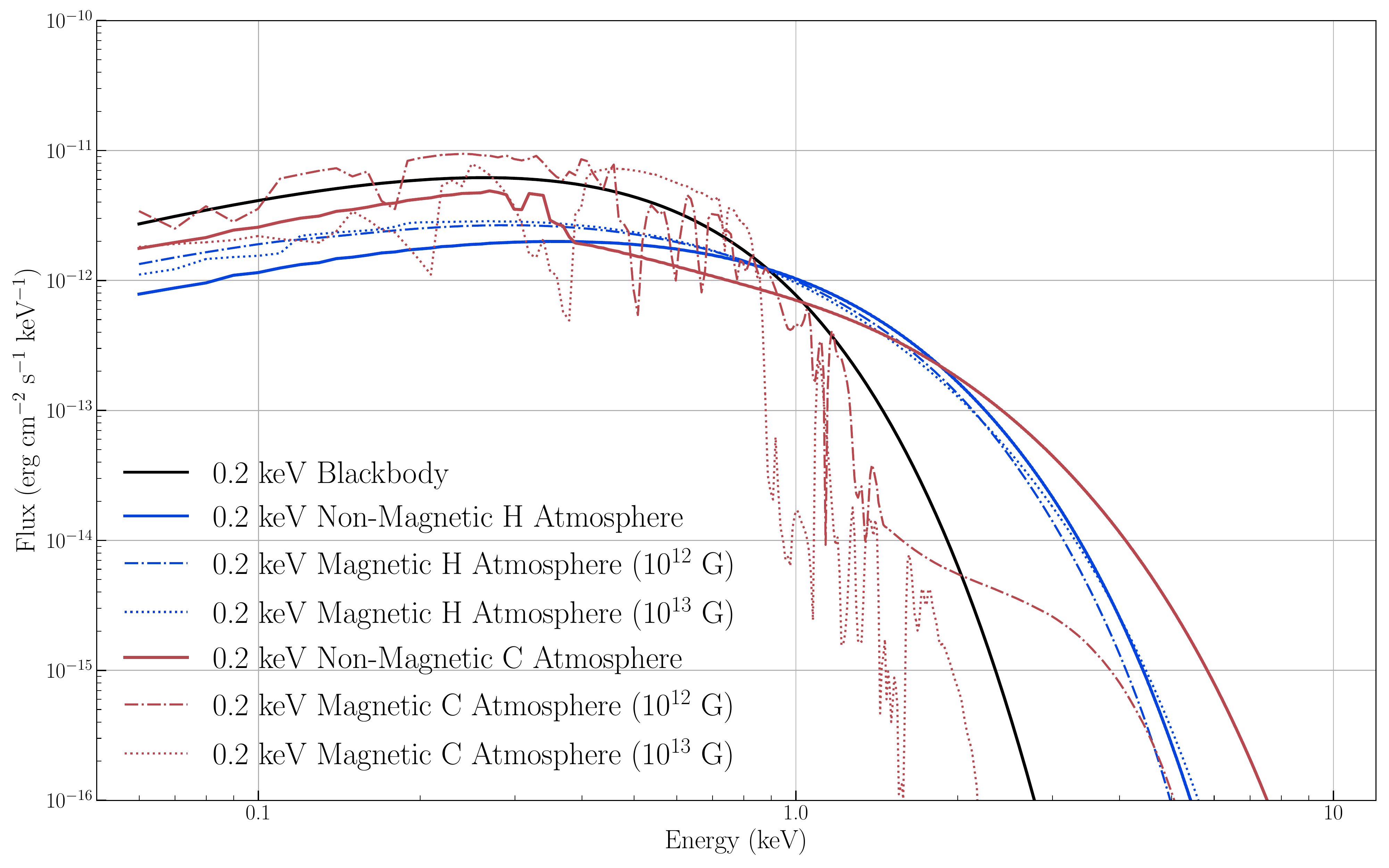}

\caption
{
\label{fig:carb_vs_bb}
The spectrum of a NS with a carbon atmosphere compared to a blackbody spectrum, and also a hydrogen atmosphere.  
The non-magnetic carbon atmosphere and non-magnetic hydrogen atmosphere models plotted here are the {\tt carbatm} and {\tt hatm} table models implemented in XSPEC.  There are narrow spectral features at low energies in the non-magnetic carbon atmosphere model that are not shown here because their width is much smaller than the energy binning used to construct the XSPEC table model. 
Their width is also much smaller than the \xmm\ energy resolution. 
Magnetic hydrogen and carbon models are shown for comparison, and they are computed from the {\tt nsmaxg} model in XSPEC
 \citep{mor07,ho08,ho14}.
All models are for a 12~km radius, $1.4\,M_{\odot}$ NS at a distance of 1~kpc, with a $0.2$~keV effective temperature..
}
\end{figure*}

Figure \ref{fig:carb_vs_bb} shows a comparison of NS X-ray spectra created by carbon atmospheres, hydrogen atmospheres, and a blackbody, all at the same 0.2 keV effective temperature $T_{\rm eff}$.  
These spectra are calculated for a 1.4\,$\msun$  neutron star with a 12~km radius, at a distance of 1~kpc.
Both magnetic and non-magnetic models are shown.
In this paper we will be considering non-magnetic carbon atmosphere models because these are the models that have been applied to CCOs. 

At a given effective temperature, the blackbody flux is greater at lower energies and the non-magnetic carbon atmosphere flux is greater at higher energies.
The energy at which the blackbody spectrum and carbon atmosphere spectrum have the same specific intensity is set by the ratio of photon absorption and scattering processes in the NS atmosphere.
Photon scattering processes dominate in the lower energy part of the spectrum and processes where photons are truly destroyed (i.e., bound-bound and bound-free transitions) dominate in the higher energy part of the spectrum.
When these processes are approximately equal, the NS atmosphere flux is approximately equal to the flux of a blackbody with the same effective temperature.
For more details on the non-magnetic carbon atmosphere model used in this work see \cite{sul14} and \cite{sul16}. 

Because the non-magnetic carbon atmosphere spectrum is harder than a blackbody with the same effective temperature, a blackbody model fitted to a given NS spectrum will have a higher effective temperature and smaller emitting area than a non-magnetic carbon atmosphere model fitted to the same data.
An effective temperature calculated from a fit to a non-magnetic carbon atmosphere model will be a factor $f$ lower than the blackbody temperature $T_{BB} = f T_{\rm carbon}$, where $f$ is usually in the range 1.5--3, and the implied radius of the UTCA model is roughly a factor of $f^{2}$ larger: $R_{\rm carbon} \sim f^{2} R_{BB}$.

The non-magnetic carbon atmosphere spectra are harder than the non-magnetic hydrogen atmosphere spectra, which are harder than the blackbody spectra, so the areas of the implied emitting regions decrease significantly from carbon atmosphere, to hydrogen atmosphere, to blackbody spectral models.
Though the available magnetic carbon atmosphere models which have $B=10^{12-13}$~G are probably not applicable to CCOs, we note that they have spectral shapes and implied emitting areas similar to blackbodies.
Even if they were applicable to CCOs, their radiation would be coming from only a small fraction of the NS surface, therefore requiring non-uniform surface temperature.

In practice, the X-ray spectra of CCOs are often consistent with carbon atmosphere, hydrogen atmosphere and blackbody spectra.
Distinguishing spectral features that would be produced by non-magnetic carbon atmospheres cannot be resolved with currently available data.  
Conclusions that some CCOs have UTCAs rely on additional probabilistic reasoning and assumptions that we will review next.

\subsection{Observational Evidence of Carbon Atmospheres}
\cite{ho09} proposed that the CCO in the Cas A SNR has a single-temperature carbon atmosphere, with the X-ray emission coming from the whole surface of the NS.
Subsequently, \cite{klo13,klo16} concluded that two candidate CCOs, in the G353.6$-$0.7 and G15.9+0.2 supernova remnants, have UTCAs.
\cite{sul17} presented calculations arguing that the probability of uniform-temperature surface for the CCO in G353.6$-$0.7 \ is $91.8 \%$. 
Using the same probabilistic reasoning, \cite{dor18} suggested that a UTCA covering the entire surface of the CCO in G330+1.0 is more plausible than the alternative, small hot-spot scenario.
These calculations of probability employed a critical assumption: that the locations of the strong magnetic fields required to produce small hot-spots, such as the hots spots known to exist in the three pulsed CCOs, are completely uncorrelated with the NS rotation axis.  That is, the distribution of hot-spots on the surface is random.  
But there is already evidence that the locations of the hot-spots on \pupa, \twelveohseven, and \kesseventynine\ are not random \citep{alf22,bog14}.  In this paper, we will review this evidence, and find that an assumed random distribution of hot-spots is also contradicted by the very low upper limits on pulsations from the CCOs in the \velajrsnr\ and \gthreefortysevensnr\ SNRs.

The UTCA models are otherwise attractive in that they can fit the phase-averaged X-ray spectra with reasonable inferred values of the NS radius, while models of other atmospheric compositions and blackbody models give radii that are much smaller than a NS radius.  The reasonable NS radii implied by the non-magnetic carbon atmosphere model might be considered strong evidence if the model predicted reasonable NS radii in only a small region of the model's NS radius$-$distance parameter space, i.e., if the model were easily falsifiable. But, we will show in Section \ref{section:spec_analysis} that the carbon atmosphere models allow reasonable NS radii for a large range of distances, i.e., their predictive power is weak.

Finally, the small areas implied by the blackbody models do not pose any theoretical difficulty.
It is known that at least the three pulsed CCOs do not have uniform-temperature surfaces; they \emph{do} in fact have heated surface regions that are significantly smaller than the NS radius \citep{got13}.  In summary, there is ample reason to critically evaluate the general applicability of the UTCA model to CCOs.

\subsection{Organization of the remainder of this Paper}
We use relevant archival data on CCOs from \xmm\ and Chandra.
In Section \ref{section:spec_analysis} we fit the X-ray spectrum of every CCO to a non-magnetic UTCA model.
We demonstrate that some CCOs without detected X-ray pulsations do nevertheless have small hot-spots, because they are at distances too close for the X-ray emission to originate from the whole NS surface even with a carbon atmosphere.
We also show that the evidence for uniform-temperature surfaces on some other CCOs is not strong because there are large ranges of distances that yield reasonable NS radii.
In Section \ref{section:timing} we calculate updated limits on pulsed fraction for the non-pulsing CCOs.
In Section \ref{section:spot_modeling} we quantify how correlated CCO hot-spot locations must be with the NS rotation axis.
In Section \ref{section:discusion}  we compare our results with previous studies, and discuss the physics of the localized thermally emitting regions.

\section{Spectral Analysis}
\label{section:spec_analysis}

\subsection{Data Reduction and the {\tt carbatm} Model }
We perform the X-ray spectral analyses using XSPEC version 12.12.0, HEASOFT version 6.29, \xmm \ SAS version 18.0.0, CIAO version 4.14, and CALDB version 4.9.6.
CALDB version 4.9.6 includes the latest \chandra \ ACIS contaminant model N0014, which is  important for accurately modeling the CCO in Cas A.
We selected one or more of the highest quality observations for each CCO.
Table~\ref{tbl:cco_obs} lists the specific observations  used for the spectral analyses.

We use the non-magnetic carbon atmosphere model {\tt carbatm}, available in XSPEC.  The
{\tt carbatm} model has four parameters: the NS effective temperature, mass, radius and normalization ($T_{\rm eff}$, $M_{\rm NS}$, $R_{\rm NS}$, $K = A / D_{10 \ \rm kpc}^{2}$).
The normalization parameter $K$ is a function of the distance $D$ to the NS and the fraction $A$ of the star surface emitting the radiation.
We set $A = 1$ so that the thermal emission originates from the whole surface, making the normalization only a function of distance.
See \cite{sul14,sul16} for further details on the {\tt carbatm} model.  

We use the {\tt tbabs} model to describe the effect of intervening column density $N_{\rm H}$, and we fit the {\tt tbabs*carbatm} model to each observation, holding the NS mass fixed at $1.4\,\msun$, and searching through a range of distances and NS radii.
At at each fixed distance and radius, we allow the temperature and column density $N_{\rm H}$ to vary to fit the data. 
We record the null hypothesis probability for each fixed distance and radius.
For most CCOs, we plot these null hypothesis probability values as contours for all plausible values of the NS radius and distance (Figures~\ref{fig:vela_jr_g347_spec}--\ref{fig:g353}).  Gray shaded regions in the Figures indicate the range of independently measured distances to the SNRs.  
Table~\ref{tbl:cco_distances} lists these distance measurements, along with a brief description of the method used.
Table~\ref{tbl:carbatm_results} lists the results of fitting the spectrum of every CCO to the {\tt carbatm} model.
The sizes of the allowed regions in the NS distance-radius parameter space quantify the falsifiability of the {\tt carbatm} model.  
We plot the spectra and {\tt carbatm} model, with distance values that are good fits to the spectral data (even if different from the independent estimates), and a plausible NS radius.

\begin{deluxetable*}{lcccccc}
\tablecaption{Log of X-ray Observations for Spectral Analysis}
\tablehead{
\colhead{CCO} & \colhead{SNR} & \colhead{Date} & \colhead{Observatory} & \colhead{ObsID}  & \colhead{Exposure} & \colhead{Instr./Mode}  \\
\colhead{} & \colhead{} & \colhead{(UT)} & \colhead{} & \colhead{} & \colhead{(ks)} }
\startdata
\pupa & Puppis A & various\tablenotemark{a} & \xmm & various  & 471 & EPIC-pn/SW \\
\velajr & G266.1$-$1.2 & 2005-06-02 & \xmm & 0207300101 & 36.9 & EPIC-pn/SW \\
\velajr & G266.1$-$1.2 & 2010-11-13 & \xmm & 0652510101 & 52.7 & EPIC-pn/SW \\ 
\twelveohseven & PKS 1209$-$51/52 &  2017-06-22  & \xmm &  0800960201   &  34.8  & EPIC-pn/SW   \\
\twelveohseven & PKS 1209$-$51/52 &  2017-06-23  & \xmm &  0800960301   &  22.2  & EPIC-pn/SW   \\
\twelveohseven & PKS 1209$-$51/52 &  2017-06-24  & \xmm &  0800960401   &  24.1  & EPIC-pn/SW   \\
\twelveohseven & PKS 1209$-$51/52 &  2017-07-03  & \xmm &  0800960501   &  25.0  & EPIC-pn/SW   \\
\twelveohseven & PKS 1209$-$51/52 &  2017-08-10  & \xmm &  0800960601   &  21.3  & EPIC-pn/SW   \\
\gthreethirty & G330.2+1.0 & 2015-03-08 & \xmm & 0742050101 & 122.0 & EPIC-pn/FW \\
\gthreefortyseven & G347.3$-$0.5 & 2013-08-24 & \xmm & 0722190101 & 94.9 & EPIC-pn/SW \\
\gthreefortyseven & G347.3$-$0.5 & 2014-03-02 & \xmm & 0740830201 & 77.2 & EPIC-pn/SW \\
\gthreefifty & G350.1$-$0.3 & 2007-02-23 & \xmm & 0402040101 & 29.3 & EPIC-pn/FW \\
\kesseventynine &  Kes~79 &   various\tablenotemark{a}  & \xmm &  various   &  498.8  & EPIC-pn/SW \\
\casa & Cas~A & 2006-10-19 & \chandra & 6690 & 61.6 & ACIS-S/TE \\
\casa & Cas~A & 2012-05-05 & \chandra & 13783 & 63.4 & ACIS-S/TE \\
\casa & Cas~A & 2015-04-27 & \chandra & 16946 & 68.1 & ACIS-S/TE \\
\hline
\gthreefiftythree & G353.6$-$0.7 & 2007-03-02 & \xmm & 0405680201 & 19.5 & EPIC-pn/FW \\
\gthreefiftythree & G353.6$-$0.7 & 2014-02-24 & \xmm & 0722190201 & 87.6 & EPIC-pn/SW \\
\gfifteen & G15.9+0.2 & 2015-07-30 & \chandra & 16766 & 92.0 & ACIS-S/TE 
\enddata
\tablecomments{
Above the line are the eight confirmed CCOs; below the line are two candidates. 
}
\tablenotetext{a}{ The spectral data reduction procedures for \pupa\ and \kesseventynine\ are the same as described in \citet{alf22} and \cite{bog14}, respectively.}
\label{tbl:cco_obs}
\end{deluxetable*}

\begin{deluxetable*}{lcccc}
\tablecaption{Independent SNR Distance Measurements}
\tablehead{
\colhead{CCO} & \colhead{SNR} & \colhead{Distance (kpc)} & \colhead{Method} & \colhead{Ref.} 
}
\startdata
RX J0822.0$-$4300 & Puppis~A & 1.3 $\pm$ 0.3 & \ion{H}{1} Velocity & 1 \\
CXOU J085201.4$-$461753 & G266.1$-$1.2 (Vela Jr.) & 0.5$-$1.0 & X-ray Expansion and & 2 \\
 &  &  &  Molecular Cloud Association &  \\
1E 1207.4$-$5209 & PKS 1209$-$51/52 & $2.1^{+1.8}_{-0.8}$ & \ion{H}{1} Velocity & 3 \\
CXOU J160103.1$-$513353 & G330.2+1.0 & $>$4.9 & \ion{H}{1} Velocity & 4 \\
1WGA J1713.4$-$3949 & G347.3$-$0.5 & 1.3 $\pm$ 0.4 & \ion{H}{1} Velocity & 5 \\
XMMU J172054.5$-$372652 & G350.1$-$0.3 & $\sim$ 4.5 & \ion{H}{1} and $^{12}$CO Velocity & 6 \\
CXOU J185238.6+004020 & Kes~79 & 6.5$-$7.5 & \ion{H}{1} Velocity & 7 \\
CXOU J232327.9+584842 & Cas~A & 3.33 $\pm$ 0.10 & Optical Expansion & 8 \\
\hline 
XMMU J173203.3$-$344518 & G353.6$-$0.7 & 3.2 $\pm$ 0.8 & Association with \ion{H}{2} region G353.42$-$0.37 & 9 \\
CXOU J181852.0$-$150213 & G15.9+0.2 & 7$-$16 & \ion{H}{1} Velocity & 10 \\
\enddata
\tablecomments{Above the line are eight well-established CCOs; below the line are two candidates.}
\tablerefs{
(1) \citealt{rey17};
(2) \citealt{all15};
(3) \citealt{gia00};
(4) \citealt{mcc01};
(5) \citealt{cas04};
(6) \citealt{gae08};
(7) \citealt{gia09};
(8) \citealt{ala14};
(9) \citealt{tia08};   
(10) \citealt{tia19}.
}
\label{tbl:cco_distances}
\end{deluxetable*}

\begin{deluxetable*}{lccccccccc}
\tablecaption{Results of Spectral Fits to the {\tt tbabs*carbatm} Model}
\tablehead{
\colhead{SNR} & \colhead{Date} & \colhead{$N_{\rm H}$} &  \colhead{$T_{\rm eff}$} & \colhead{$D$ (fixed)}  & \colhead{$R_{\rm NS}$ (fixed)} & \colhead{$M_{\rm NS}$ (fixed)} & \colhead{$\chi^{2}_{\nu}$ (d.o.f.)} & \colhead{$P_{\rm Null \ Hyp.}$}\\
\colhead{} & \colhead{(UT)} & \colhead{($10^{22}$ cm$^{-2}$)} &  \colhead{($10^6$ K)} & \colhead{(kpc)}  & \colhead{(km)} & \colhead{($\msun$)} & \colhead{} & \colhead{} }
\startdata
Puppis~A & various & $0.348^{+0.004}_{-0.004}$ & $1.531^{+0.001}_{-0.002}$ & 1.3 & 12.0 & 1.4 & 1.06 (208) & 0.250  \\
Vela Jr. & 2005-06-02 & $0.470^{+0.012}_{-0.012}$ & $1.533^{+0.005}_{-0.005}$ & 2.5 & 13.0 & 1.4 & 1.13 (73) & 0.209  \\
Vela Jr.  & 2010-11-13 & $0.491^{+0.011}_{-0.011}$ & $1.536^{+0.004}_{-0.004}$ & 2.5 & 13.0 & 1.4  & 1.09 (80) & 0.265 \\
PKS 1209$-$51/52 & various & $0.099^{+0.004}_{-0.006}$ & $1.484^{+0.005}_{-0.005}$ & 2.0 & 12.0 & 1.4 & 1.01 (145) & 0.463  \\
G330.2$+$1.0 & 2015-03-08 & $3.105^{+0.147}_{-0.151}$ & $1.801^{+0.023}_{-0.021}$ & 6.0 & 12.0 & 1.4 & 1.19 (80) & 0.115   \\
G347.3$-$0.5 & 2013-08-24 &  $0.527^{+0.004}_{-0.004}$ &   $2.102^{+0.003}_{-0.003}$ & 2.3 & 9.0 & 1.4 & 1.24 (93) & 0.057  \\
G347.3$-$0.5 & 2014-03-02 & $0.518^{+0.005}_{-0.005}$ & $2.045^{+0.003}_{-0.004}$ & 2.6 & 10.0 & 1.4 & 1.09 (92) & 0.260 \\
G350.1$-$0.3 & 2007-02-23 & $3.575^{+0.216}_{-0.233}$ & $2.206^{+0.034}_{-0.033}$ & 6.0 & 12.0 & 1.4 & 1.03 (38) & 0.416 \\
 Kes~79 & various & $1.737^{+0.050}_{-0.052}$ & $1.841^{+0.011}_{-0.011}$ & 7.0 & 12.0 & 1.4 & 0.97 (112) & 0.577  \\
Cas~A & 2006-10-19 & $1.453^{+0.044}_{-0.043}$ & $1.785^{+0.011}_{-0.011}$ & 3.33 & 12.0 & 1.4 & 1.109 (53)   & 0.272 \\
Cas~A & 2012-05-05 & $1.467^{+0.051}_{-0.050}$ & $1.779^{+0.012}_{-0.012}$ & 3.33 & 12.0 & 1.4 & 0.900 (50)  & 0.675 \\
Cas~A & 2015-04-27 & $1.395^{+0.054}_{-0.052}$ & $1.758^{+0.011}_{-0.012}$ & 3.33 & 12.0 & 1.4 & 0.872 (47)  & 0.719 \\
\hline
G353.6$-$0.7 & 2007-03-21 & $1.771^{+0.046}_{-0.048}$ & $2.183^{+0.013}_{-0.013}$ & 3.2 & 12.0 & 1.4 & 1.16 (70) & 0.169  \\
G353.6$-$0.7 & 2014-02-24\tablenotemark{a} & 1.898 & 0.191 & 3.2 & 12.0 & 1.4 & 2.23 (92) & 0.000 \\
G15.9+0.2 & 2015-07-30 & $4.109^{+0.368}_{-0.410}$ & $1.702^{+0.044}_{-0.041}$ & 8.5 & 12.0 & 1.4 & 0.61 (26) & 0.941
\enddata
\tablecomments{
The parameters listed here correspond to the spectral models shown in Figures \ref{fig:vela_jr_g347_spec}--\ref{fig:g353}.
Distances were chosen primarily to achieve an acceptable fit to the spectral data, a reasonable NS radius, and consistency with the independently measured distances in Table \ref{tbl:cco_distances}, if possible. }
 \tablenotetext{a}{This \xmm\ observation is incompatible with the {\tt carbatm} model with a 1.4 $\msun$ NS; we have listed the best fit parameters here even though the fit is poor.} 
\label{tbl:carbatm_results}
\end{deluxetable*}

\subsection{Vela Jr. and G347.3$-$0.5}
\label{section:vela_jr_g347}
Figure \ref{fig:vela_jr_g347_spec} shows that the spectra of  CCOs in \velajrsnr \ and \gthreefortysevensnr\ are consistent with the {\tt carbatm} model.
However, the independent distance estimates imply that both NSs are too close to have UTCAs \ with reasonable implied NS radii.
The closer distances of these two CCOs implies that they must actually have small hots spots and not uniform-temperature surfaces.

\subsection{The Three Pulsing CCOs}
The X-ray pulses from the CCOs in \pupasnr, \kesseventyninesnr, and \twelveohsevensnr \  indicate they \emph{do not} have uniform-temperature surfaces.
Nevertheless, Figure~\ref{fig:pupa_kes_1207_spec} shows that the X-ray spectra of these three CCOs are all consistent with the {\tt carbatm} model.
The {\tt carbatm} model yields reasonable NS radii, and is consistent with the independent distance estimates.
{We have not included any spectral lines in the model of the CCO in \kesseventyninesnr. There is a hint of one or more spectral features in the fit residuals, though the fit is formally acceptable without the inclusion of any line features.  Our model of the CCO in Puppis A includes a Gaussian absorption line at 0.46 keV, though an emission line around 0.7 keV would also fit the data.  An emission or absorption feature was required in the two-blackbody fit to the \pupa\ spectrum \citep{got09,del12}.

Fitting a carbon atmosphere model to the spectrum of \twelveohseven \ required the inclusion of equally spaced absorption features at 0.7, 1.4, 2.1 and 2.8 keV.
The features were known to be present in two-blackbody spectral fits, and are required to fit the UTCA \ model as well.
{The 0.7 keV absorption feature is due to classical electron cyclotron absorption, while the harmonics at 1.4, 2.1 and 2.8 keV are  due to quantum oscillations in the atmospheric free-free opacity and also relativistic effects \citep{sul10,sul12}.
We model the absorption features as Gaussian lines with their relative energy spacings held constant, and their strengths and widths allowed to vary to fit the data.

\subsection{\gthreethirtysnr, \gthreefiftysnr, \gfifteensnr}
Figure \ref{fig:g330_g350_g15_spec} shows the fits of the CCOs in \gthreethirtysnr, \gthreefiftysnr, and \gfifteensnr\ to the {\tt carbatm} model.
We find that the {\tt carbatm} model is consistent with a large range of NS radii and distances.
The large uncertainties in the CCO distances, and the low quality of the spectral data, make it difficult to use these to either support or falsify the {\tt carbatm} model.

\subsection{\casasnr}
The CCO in Cas~A is the first NS proposed to have a carbon atmosphere \citep{ho09}.
Additionally, \cite{hei10} reported that its surface temperature cooled by $4\%$ between the years 2000 and 2009, the first reported direct detection of NS cooling.
This apparent cooling is only evident under the assumption that the {\tt carbatm} model is correct; the data are equally well described by a hydrogen atmosphere model, where the emission is coming from small hot-spots that are shrinking in size at constant temperature \citep{pos18,pos22}.

Due to the brightness of the Cas~A SNR, observations performed in \chandra \ observing modes with small frame integration times are required to limit spectral distortions due to photon ``pile-up''.
\chandra \ ObsIDs 6690, 13783, 16946, 17639, 22426, and 23248 were performed in faint telemetry format, with 0.34~s frame times that minimize pile-up.  For a detailed analysis of these observations see \citet{pav09}, \citet{pos13}, \citet{pos18}, and \citet{pos22}. 

\cite{pos18} analyzed  ObsIDs 6690, 13783, 16946, and 17639, and  found no statistically significant evidence of NS cooling, with conservative $3\sigma$ upper limits of  $< 3.3\%$ and $< 2.4\%$ per 10 years, for variable and fixed $N_{\rm H}$ values, respectively.  \cite{pos22} analyzed all six of these \chandra observations, and found that they could be equally well described by non-magnetic hydrogen or carbon atmosphere models. 
\cite{pos22} also found an apparent increase in the cooling rate over the last five years of observations, with a rate of  $-2.3 \pm 0.4 \%$ per 10 years ($N_{\rm H}$ allowed to vary independently between observations).
\cite{ho21} presented an analysis of all ``graded'' \chandra \ ACIS data, with $\approx 3$~s frame times and larger pile-up fractions ranging from $6\%$ to $13\%$.
They found that the graded \chandra \ ACIS data are also consistent with a non-magnetic carbon atmosphere, with an implied cooling rate of $2.8 \pm 0.3 \%$ per 10 years (one-sigma error, $N_{\rm H}$ allowed to vary independently between observations).

We analyzed \chandra \ ObsIDs 6690, 13783, and 16946 using the same source and background regions as \cite{pos18,pos22}.
We binned the spectra with a signal-to-noise ratio of at least 10 and used the latest \chandra \ ACIS contaminant model N0014. 
We fit the {\tt carbatm} model to them simultaneously, allowing $T_{\rm eff}$ and $N_{\rm H}$ to vary independently, and keeping $M_{\rm NS}$, $R_{\rm NS}$, and $D$ fixed at equal values for each observation.
Figure~\ref{fig:cas_a_spec} shows the spectra of the Cas~A CCO obtained during these observations.
The primary difference between our analysis here and previous works is that we are searching over both $R_{\rm NS}$ and $D$ parameter space in order to see how falsifiable the {\tt carbatm} model is based on spectra alone. 
We can see in Figure~\ref{fig:cas_a_spec} that the precisely measured distance to Cas A ($D = 3.33 \pm 0.10$ kpc), is not significantly constrained by the model; the highest quality available data can still accommodate a significant range of distances and reasonable NS radii.
We list the spectral parameters corresponding to a $M_{\rm NS} =1.4 \msun$ and $R_{\rm NS} = 12$~km  in Table \ref{tbl:carbatm_results}.
\cite{pos22} found slightly higher values $T_{\rm eff}$ implied by the {\tt carbatm} model.
We have confirmed that these differences in $T_{\rm eff}$ are due to the different assumed NS mass and radius, since the more compact NS parameters assumed by 
\cite{pos22} imply a larger gravitational redshift.

\subsection{G353.6$-$0.7}
\label{section:g353_spec}
Figure \ref{fig:g353} shows two \xmm\ spectra of the CCO in \gthreefiftythreesnr.  
The earlier  spectrum was fitted by \cite{klo13}, \cite{klo15} and \cite{sul17} to a UTCA.  
But the more recent, longer observation is inconsistent with the {\tt carbatm} model.
Figure \ref{fig:g353} also shows that it has an acceptable fit to a two blackbody model.
The flux of the hotter component is dominant over the flux from the cooler component.
The cooler component likely originates from the entire NS surface, although its temperature and area are not well constrained, so we have fixed the temperature $kT_{\rm warm} = 0.08$ keV in this analysis.
In the absence of detected X-ray pulsations, and without a precise distance estimate, the data do not strongly constrain the temperature and emitting area of the cooler spectral component. 

The X-ray emission from the CCO in \gthreefiftythreesnr \ has likely been absorbed and scattered by dust along our line of sight \citep{hal10b,lan22}.
The net effect on the observed spectrum would depend on the location of the dust scattering screen(s) and the grain sizes, both of which are unknown.
We attempted to account for this dust with the {\tt xscat} model available in XSPEC.
We found that, assuming a 3.2 kpc distance, the CCO spectrum is still inconsistent with any UTCA model, for a wide range of NS masses and radii ($0.6-2.0~\msun$, $8-15$~km), with all reduced chi-square values $\chi_{\nu}^{2} \geq 1.4$.

Recently, \cite{Doroshenko2022} concluded that the CCO in \gthreefiftythreesnr \ has a very small mass, $M = 0.77_{-0.17}^{+0.20} \ \msun$.
 But, this result depends on \emph{all} of the following necessary but insufficient conditions:
1) the CCO has a UTCA, 2) the star associated with Gaia EDR3 source 5975119332093959552 is located inside the G353.6$-$0.7 SNR, and 3) the distance to this star, and therefore also G353.6$-$0.7, is 2.5 kpc. 
This final assumption about the distance to G353.6$-$0.7 is critical to their conclusion that the CCO mass is much less than $1.4 \ \msun$.
Extended Data Figure 1 of \citet{Doroshenko2022} indicates that, if the CCO is actually located at the 3.2 kpc distance favored by \cite{tia08}, then the $1\sigma$ uncertainties on the NS mass would still be consistent with a $\approx 1.4 \ \msun$ NS.

\cite{Doroshenko2022} did not analyze the longest, highest quality \xmm \ observation shown in the bottom two panels of Figure \ref{fig:g353}. 
We have found this observation is inconsistent with a UTCA model of the CCO, at a fixed 3.2 kpc distance, for any NS mass in the range from $0.6-2.0~\msun$, with or without attempting to model the dust scattering. 
Even allowing for a 2.5 kpc distance and a NS mass $M = 0.77~\msun$, we find a null hypothesis probability of 0.03, still worse than the two-blackbody model shown in the bottom panel of Figure \ref{fig:g353}, which does not require a small NS mass at either distance.

\begin{figure*}
\centering
\includegraphics[width=1.0\linewidth]{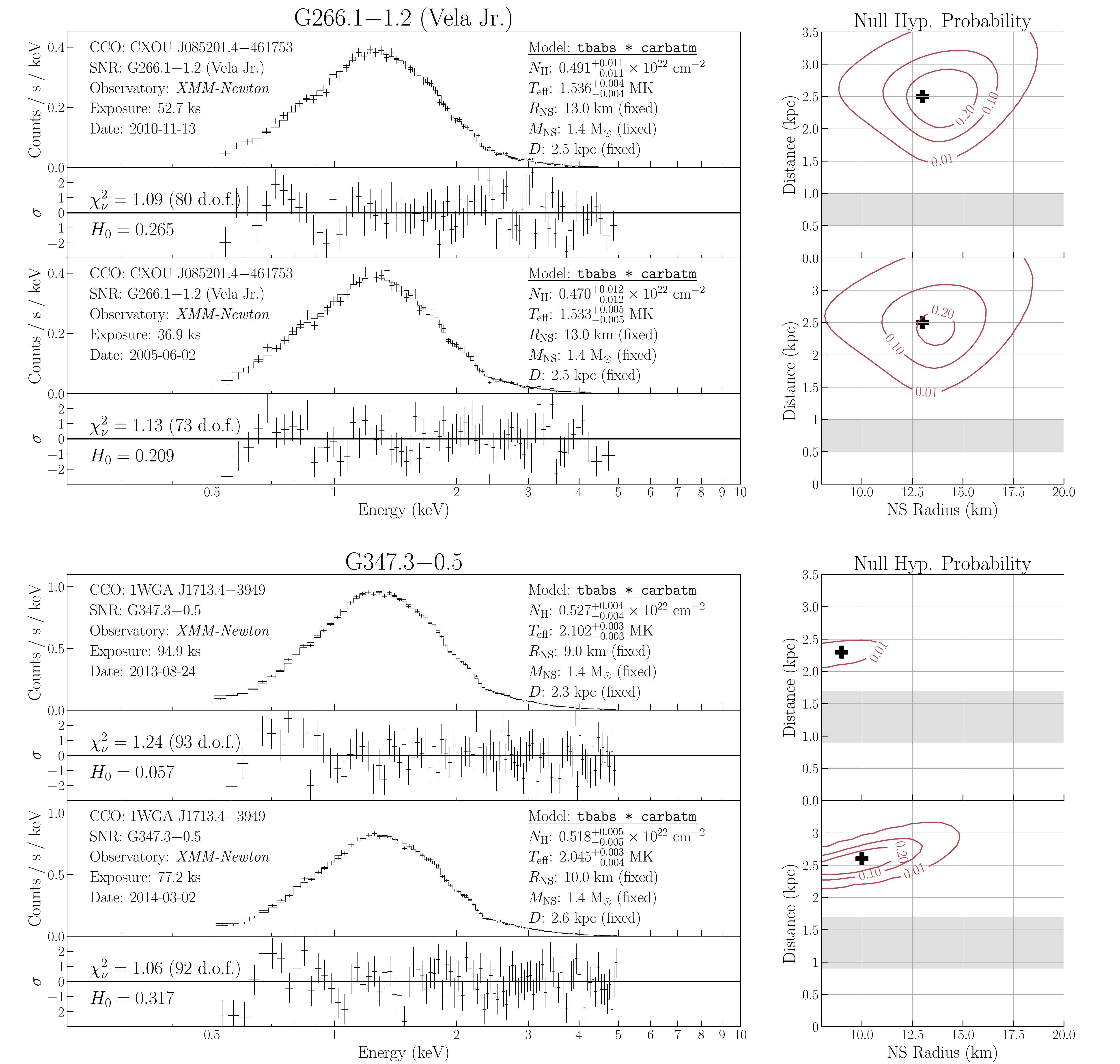}
\caption
{
\label{fig:vela_jr_g347_spec}
Here we demonstrate that the CCOs in \velajrsnr \ and \gthreefortysevensnr \ do not have uniform-temperature surfaces covered by carbon atmospheres.  
On the left we have plotted the spectra from the two longest \xmm \ observations of both CCOs.
The plus signs in the null hypothesis probability plots (right) indicate the NS radii and distance parameters in the corresponding spectral models (left).
The value of the corresponding null hypothesis probability $H_{0}$ is indicated in the lower left of each residual plot.  
The spectra are consistent with the UTCA models, with reasonable values of the NS radius.
However, the independently measured distances, shaded in gray, are too close to allow the possibility that the emission originates from the whole NS surface.  The observed X-rays must instead originate from smaller, localized regions on the NS surface.
}
\end{figure*}

\begin{figure*}
\centering
\includegraphics[width=0.8\linewidth]{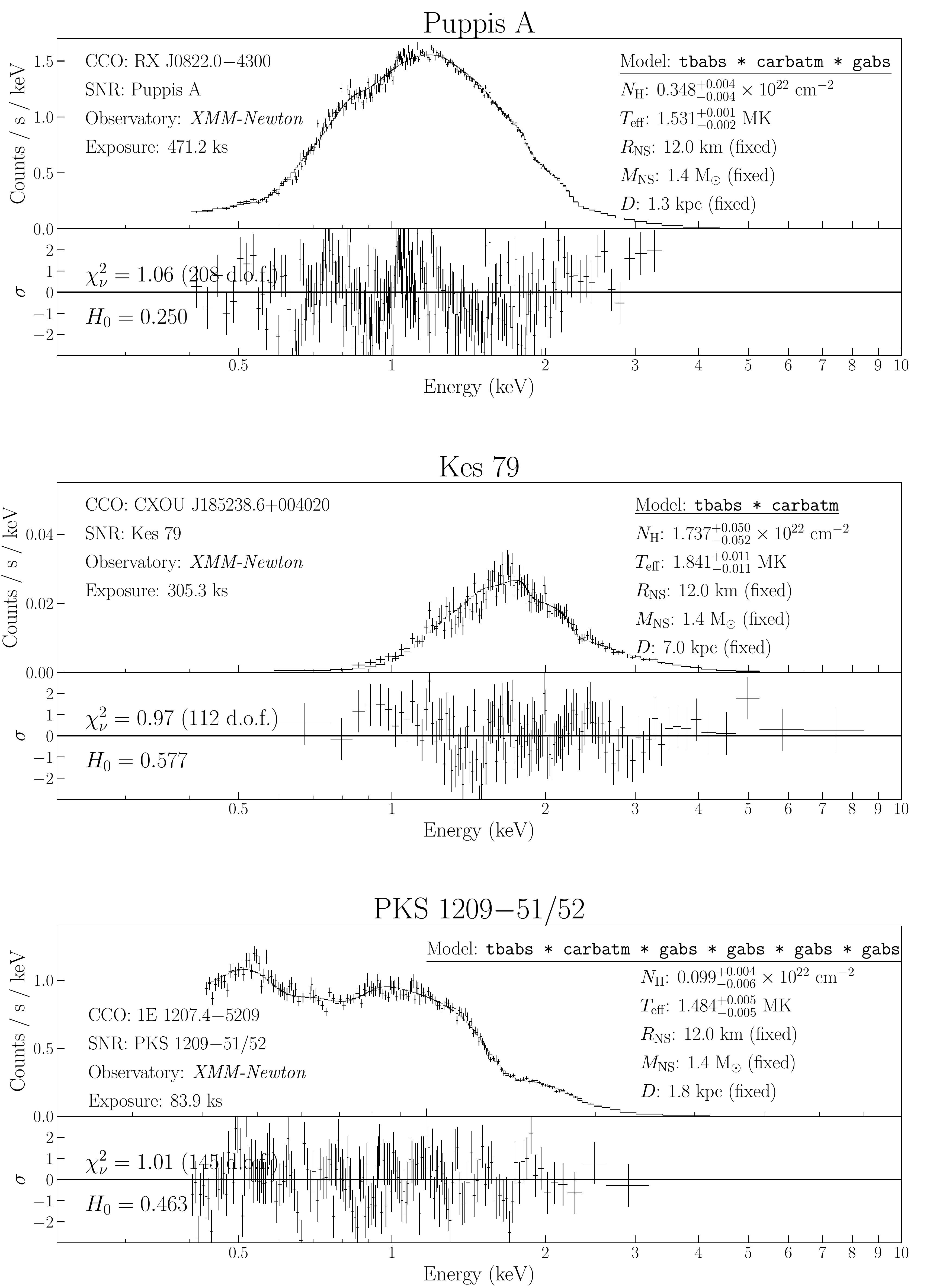}
\caption{
\label{fig:pupa_kes_1207_spec}
Here we demonstrate that the CCOs with measured X-ray pulsations can nevertheless be well described by UTCA models, with reasonable values of the NS radii and distances resulting.
It is known that the X-ray emission actually originates from hot localized regions on the surfaces of these CCOs, since these localized regions are what produce the measured X-ray pulsations.
That the spectra of these CCOs can be also be well described by UTCA models, weakens the argument for these models when applied to other CCOs.
A good fit to the UTCA model does not imply that the NS actually has a uniform surface temperature.}
\end{figure*}

\begin{figure*}
\centering
\includegraphics[width=1.0\linewidth]{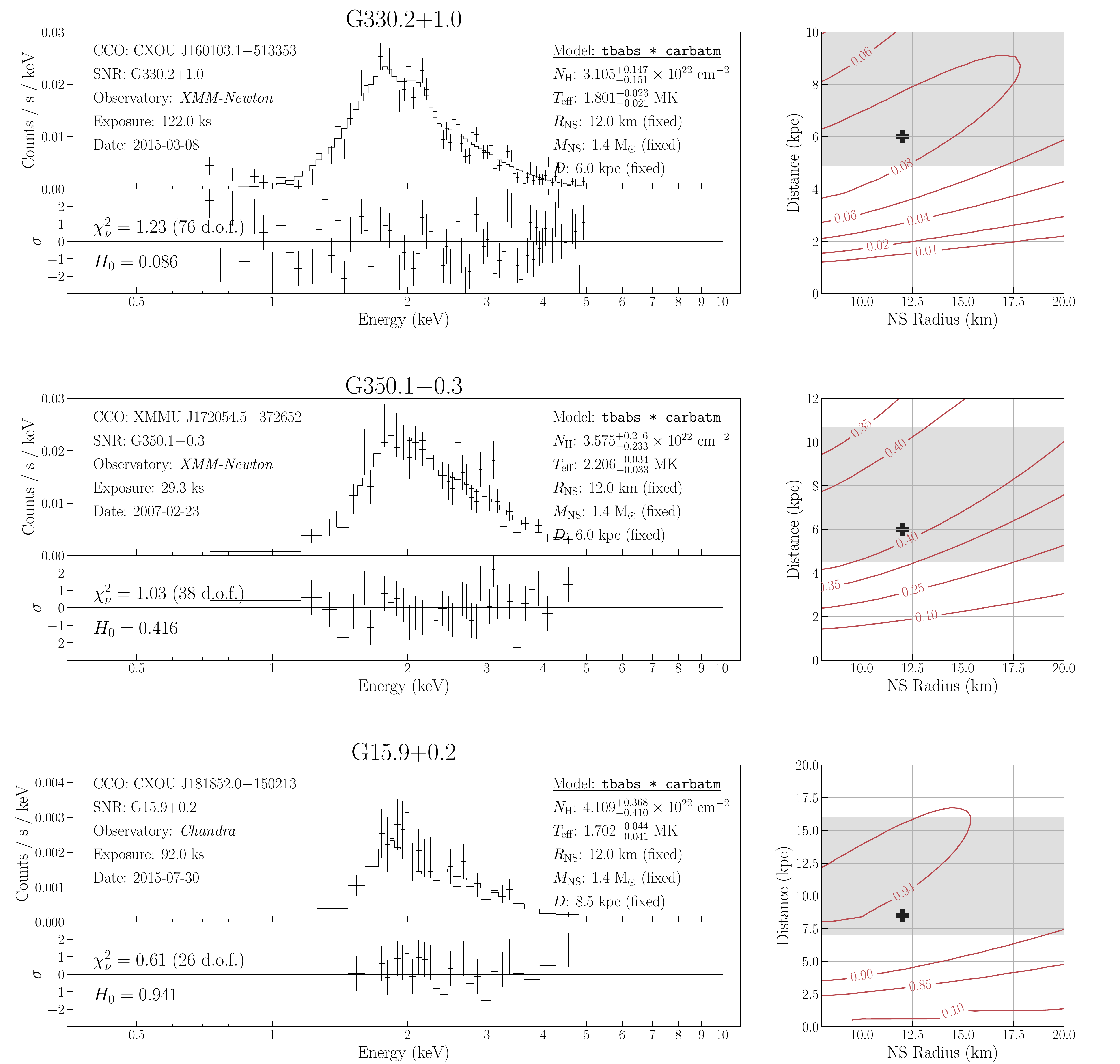}
\caption
{
\label{fig:g330_g350_g15_spec}
Here we demonstrate that when a UTCA model is a good fit to the spectrum of a CCO, there may be a large range of distances that would also work. 
For the CCOs in \gthreethirtysnr, \gthreefiftysnr, and \gfifteensnr, there are a large range of distances that would fit the model and allow for reasonable values of the NS radius.
}
\end{figure*}

\begin{figure*}
\centering
\includegraphics[width=1.0\linewidth]{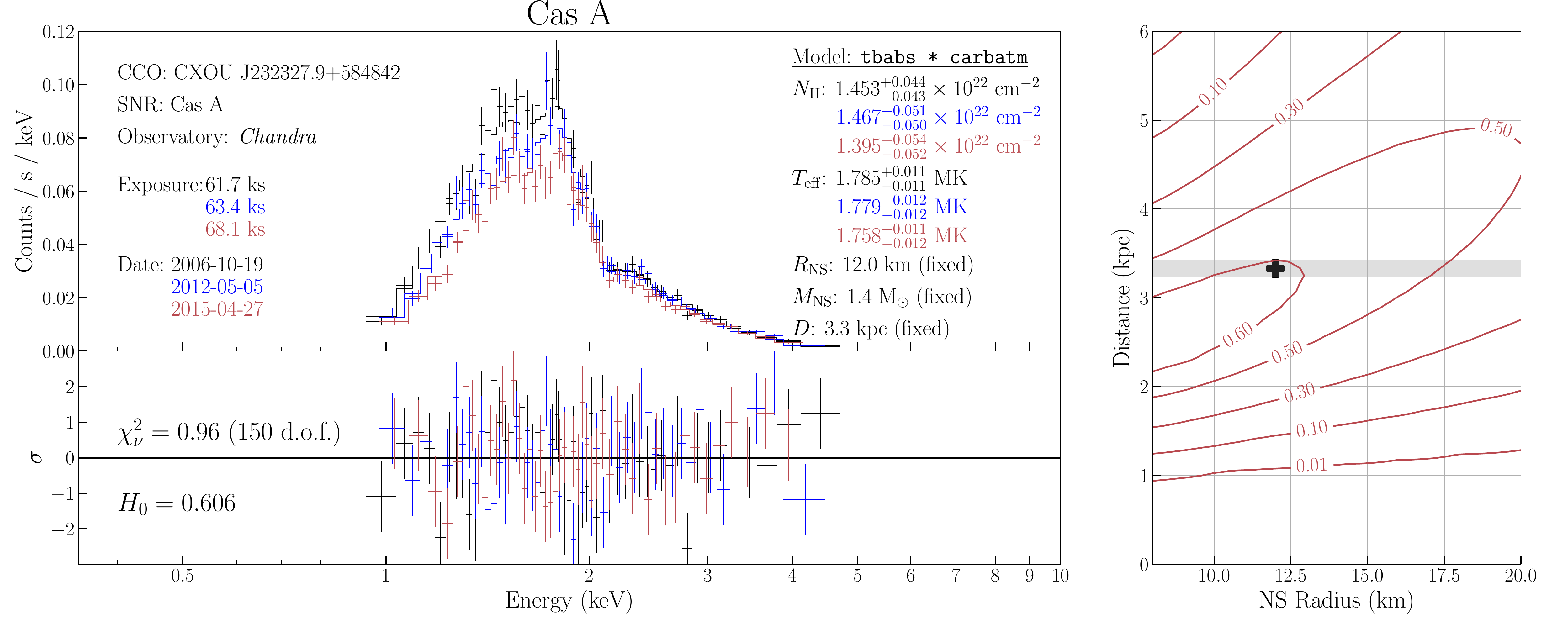}
\caption
{
\label{fig:cas_a_spec}
A simultaneous fit to the X-ray spectrum of the CCO in Cas A observed at different epochs, with the NS mass, radius and distance held fixed.
The CCO in Cas A has a precisely measured distance of $3.33 \pm 0.10$ kpc, but a closer distance estimate of $\sim 2.2$ kpc would be equally consistent with the model. 
}
\end{figure*}

\begin{figure*}
\centering
\includegraphics[width=0.75\linewidth]{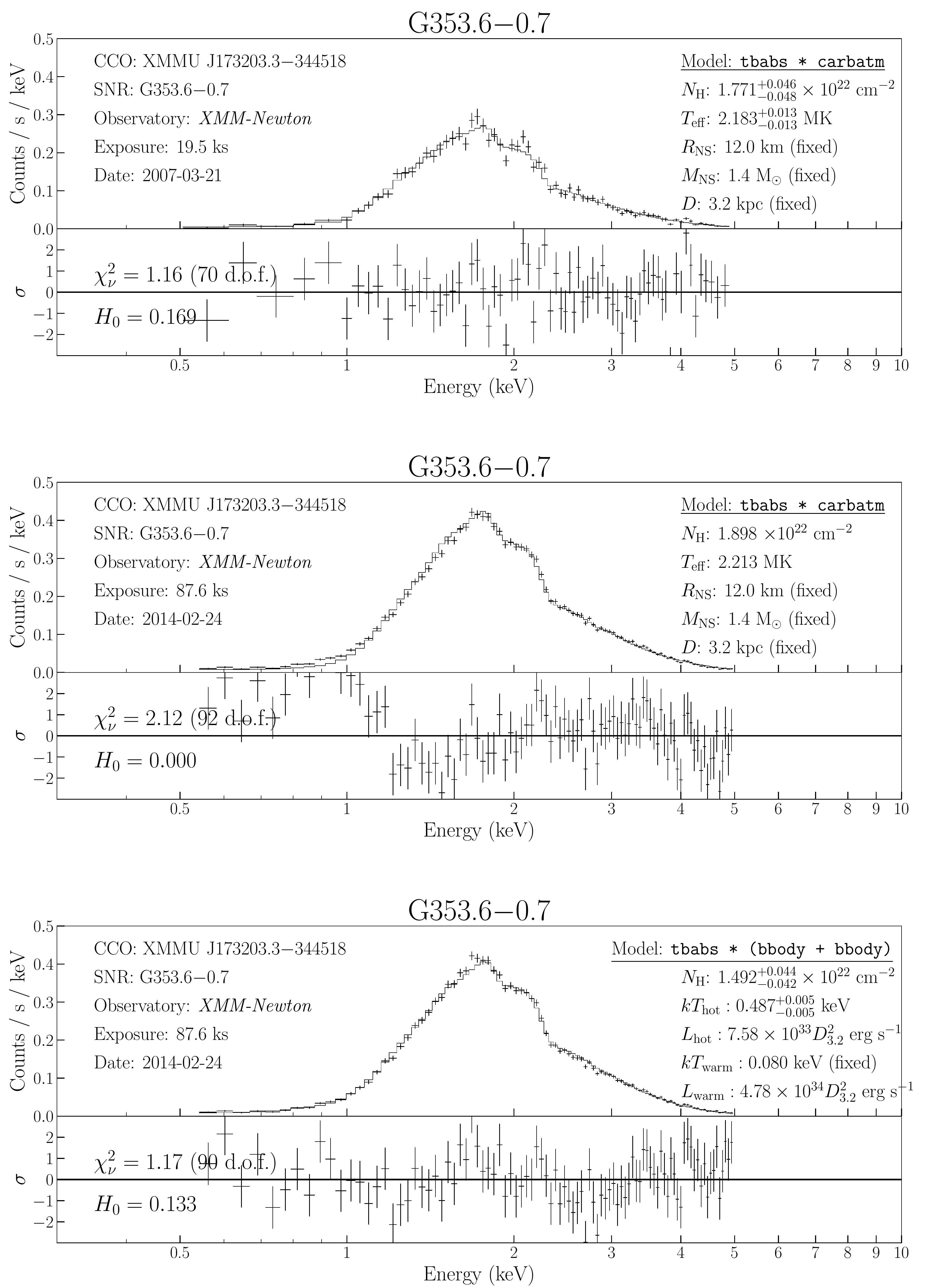}
\caption
{
\label{fig:g353}
Top: A short 19.5~ks observation of the CCO in \gthreefiftythreesnr, which is well described by a UTCA model.
Middle: The more recent 87.6~ks X-ray spectrum of \gthreefiftythree, along with a representative UTCA model. Evidently this higher-quality X-ray spectrum is inconsistent with the {\tt carbatm} model.
Bottom: The spectrum  is well described by the sum of two blackbodies, with the area of the hotter component a small fraction of the NS surface.}
\end{figure*}

\section{Upper Limits on CCO X-ray Pulsed Fractions}
\label{section:timing}

\cite{got13}, in their Table~1, reported upper limits on CCO X-ray pulsed fractions.
Here we update some of those limits with newly available timing data. 
The data sets used to search for X-ray pulsations are listed in Table~\ref{tbl:timing_obs}.

\subsection{Pulsation Search Method}
For each pulsation search we extracted events from a circular region centered on the source. 
We chose the radius of each region to maximize the signal-to-noise ratio $\mu/\sigma$ of the source photons.  Here $\mu$ is the photon count rate and $\sigma$ is the standard deviation of the photon count rate.
For each CCO, we searched for pulsations in two energy bands. One search was performed in the full energy range of the CCO spectrum, and another search was restricted to photons with energies greater than 1.5 keV.
This ensures that we do not miss pulsations from a CCO similar to \pupa, whose pulsations are almost perfectly out-of-phase in high- and low- energy bands.

\begin{deluxetable*}{lllllllll}
\tablecaption{Log of X-ray Observations Used for Timing Analysis\tablenotemark{a}}
\tabletypesize{\tiny}
\tablehead{
\colhead{SNR} 
 & \colhead{ObsID}  & \colhead{Date} 
 & \colhead{Exposure} & \colhead{Telescope/Instr./Mode} 
& \colhead{Extraction radius}  & \colhead{Counts} & \colhead{Background}  
 & \colhead{Max. $Z^{2}_{1}$} \\
 \colhead{} &   \colhead{} &  \colhead{(UT)} &  \colhead{(ks)} & 
  \colhead{} &   \colhead{} &  \colhead{} &  \colhead{(\%)} &  \colhead{} 
} 
\startdata
\textbf{G266.1$-$1.2}  \\
\hline
 & 0147750101 & 2003-05-21 & 58.0 ks & {\it XMM}/EPIC-pn/SW & $20\!^{\prime\prime}$   ($22.\!^{\prime\prime}5$) & 17922 (8278) & $16\%$ ($16\%$) & 34.3 (38.8) \\  
 & 0207300101 & 2005-06-02 & 53.9 ks & {\it XMM}/EPIC-pn/SW &
 $17.\!^{\prime\prime}5$   ($32.\!^{\prime\prime}5$) &
 19113 (9378)  & $12\%$ ($7\%$) & 35.2 (32.7)  \\
 & 0652510101 & 2010-11-13 & 84.5 ks & {\it XMM}/EPIC-pn/SW &
 $22.\!^{\prime\prime}5$   ($27.\!^{\prime\prime}5$)  &
 31390 (13604)& $15\%$ ($10\%$) & 37.1 (36.5) \\
\hline
\textbf{G330.2+1.0} \\
\hline  
& 0500300101 & 2008-03-20 & 68.4 ks  & {\it XMM}/EPIC-pn/SW
& $12.\!^{\prime\prime}5$   ($15.\!^{\prime\prime}0$) &
2210 (1437)& $47\%$ ($40\%$)   & 31.9 (30.9)\\
& 0742050101 & 2015-03-08 & 140.9 ks   & {\it XMM}/EPIC-pn/FW & 
$15.\!^{\prime\prime}0$   ($15.\!^{\prime\prime}0$) &
 4518 (3861) & $30\%$ ($27\%$) & 32.9 (33.6) \\
\hline
\textbf{G347.3$-$0.5} \\
\hline
& 0722190101 & 2013-08-24 & 138.9 ks & {\it XMM}/EPIC-pn/SW &
$37.\!^{\prime\prime}5$   ($37.\!^{\prime\prime}5$) 
& 127751 (59536) & $7\%$($8\%$) & 40.4 (35.6)   \\
& 0740830201 & 2014-03-02 & 140.8 ks & {\it XMM}/EPIC-pn/SW &
$37.\!^{\prime\prime}5$ ($37.\!^{\prime\prime}5$) & 
102660 (47948) & $7\%$($7\%$) & 34.5 (34.4) \\
\hline
\textbf{G350.1$-$0.3} \\ 
\hline
 & 14806 & 2013-05-11 & 89.7 ks & \textit{Chandra}/ACIS-S/CC &  3 pixel columns\tablenotemark{b} &  5083 ( 4716) & $15\%$($13\%$) & 35.4 (33.8)  \\
\hline
\textbf{Cas A} \\
\hline
& 1857  & 2000-10-04 & 48.9 ks & \textit{Chandra}/HRC-S & $1.\!^{\prime\prime}2$ & 1455 & $6\%$  & 32.3  \\
& 1038  & 2001-09-19 & 50.4 ks & \textit{Chandra}/HRC-S &  $1.\!^{\prime\prime}2$ & 1405 & $6\%$ & 34.5  \\
& 10227 & 2009-03-20 & 132.4 ks & \textit{Chandra}/HRC-S &  $1.\!^{\prime\prime}2$ & 3332 & $6\%$ & 36.5  \\
& 10229 & 2009-03-24 & 49.0 ks  & \textit{Chandra}/HRC-S &  $1.\!^{\prime\prime}2$ & 1243 & $6\%$ & 35.5  \\
& 10892 & 2009-03-26 & 125.7 ks  & \textit{Chandra}/HRC-S &  $1.\!^{\prime\prime}2$ & 3236 & $6\%$ & 34.6  \\
& 10228 & 2009-03-28 & 131.0 ks & \textit{Chandra}/HRC-S &  $1.\!^{\prime\prime}2$ & 3192 & $6\%$ & 33.3  \\
& 10698 & 2009-03-31 & 52.0 ks  & \textit{Chandra}/HRC-S &  $1.\!^{\prime\prime}2$ & 1352 & $6\%$ & 38.7   \\
\hline
\textbf{G353.6$-$0.7} \\ 
\hline
& 0722190201 & 2014-02-24 & 131.2 ks & {\it XMM}/EPIC-pn/SW &
$37.\!^{\prime\prime}5$   ($37.\!^{\prime\prime}5$)  &
67241 (53305) & $11\%$($10\%$) &  36.6 (32.3) \\
\hline
\enddata
\tablecomments{SW and FW indicate the Small Window mode and Full Window mode of the \xmm \ EPIC-pn detector.}
\tablenotetext{a}{List of X-ray observations analyzed to search for pulsations and place upper limits on their pulsed fractions. 
Values in parentheses refer to the searches that were restricted to photons with energies greater than 1.5 keV.}
\tablenotetext{b}{
Photons were extracted from the column centered on the CCO, and the two columns on either side.
Background rates were estimated from the two columns located three pixels away on either side of the CCO center column.
} 
\label{tbl:timing_obs}
\end{deluxetable*}

\begin{deluxetable*}{llccc}
\tablecaption{CCO Pulsed Fractions and Upper Limits}
\tabletypesize{\normalsize}
\tablehead{
\colhead{CCO} & \colhead{SNR}  & \colhead{Pulsed Fraction\tablenotemark{a}} & \colhead{Period} & \colhead{Period Ranges Searched} \\
\colhead{} & \colhead{}  & \colhead{($\%$)} & \colhead{(ms)} & \colhead{(ms)}
}
\startdata
\pupa & Puppis~A & $11$ & 112 & ... \\
CXOU J085201.4$-$461753 & G266.1$-$1.2 (Vela Jr.) &  $<5$ ($7$) & \nodata & $12-10^{7}$  \\
\twelveohseven & \twelveohsevensnr &  $9$  & 424 & \nodata \\
CXOU J160103.1$-$513353 & G330.2+1.0 & $<50$ (49) & \nodata &  $12-10^{7}$  \\
CXOU J160103.1$-$513353 & G330.2+1.0 & $<25$ (25) & \nodata & $146-10^{7}$   \\
1WGA J1713.4-3949 & G347.3$-$0.5 & $<3$ ($4$) & \nodata &  $12-10^{7}$  \\
XMMU J172054.5$-$372652 & G350.1$-$0.3 & $<19$ (19) & \nodata &  $6-10^{7}$  \\
\kesseventynine & Kes~79 & $64$  & 105 & \nodata \\
CXOU J232327.9 + 584842 & Cas A & $<11 $ & \nodata & $5-10^{7}$  \\
\hline 
XMMU J173203.3$-$344518 & G353.6$-$0.7 & $<5$ ($6$) & \nodata & $12-10^{7}$   \\
CXOU J1852.0$-$150213 & G15.9+0.2 & \nodata  & \nodata & \nodata  \\
    \hline
\enddata
\tablenotetext{a}{
Pulsed fraction upper limits ($99\%$ confidence) for the full energy range are listed, along with upper limits for searches restricted to energies greater than 1.5 keV in parentheses. 
}
\label{tbl:pf_upper_limits}
\end{deluxetable*}

We used the $Z_{1}^{2}$ statistic \citep{buc83} to search for X-ray pulsations.
The general $Z_{m}^{2}$ statistic is defined as
\begin{equation}
   Z^2_m = \frac{2}{N} \sum_{\rm k=1}^{m}
   \left[
	\left(\sum_{i=1}^{N}\cos(k\phi_i)\right)^2 + 
	\left(\sum_{i=1}^{N}\sin(k\phi_i)\right)^2
   \right]
\label{eq:z2}
\end{equation}
where $\phi_i$ is the rotational phase calculated from the arrival time and $N$ is the number of events.

All power spectra consist of noise as well as sometimes signal.  Upper limits on the X-ray pulsed fractions of NSs should be calculated from the probability distributions of power spectra including both the noise and the hypothetical signal.  \cite{gro75} calculated these probability distributions, and gave analytic formulae for the probability density $p_{n}$ that a time series of a source with intrinsic signal power $P_{s}$ will yield a measured power $P$:

\begin{equation}
p_{n}(P;P_{s}) = (P/P_s)^{(n-1)/2} \exp{[-(P+P_{s})]} I_{n-1}(2 \sqrt{P P_{s}})
\label{eq:groth}
\end{equation}
Here $I_{n-1}$ is the modified Bessel function of the first kind, and $n$ is the number of frequency bins being summed to calculate the total measured power $P$ at a particular frequency.
In this analysis, we are summing power spectra incoherently from multiple observations, so $n$ is the number of summed power spectra (we also perform a coherent search for the Cas A CCO).  
Note that there is a factor of 2 difference in the definition of $Z_{1}^{2}$ in Equation~\ref{eq:z2} and the measured pulsed power $P$ in Equation~\ref{eq:groth}: $Z_1^2 = 2P$.
We integrate the probability density $p_{n}$ to calculate $f_{n}$, the probability that the measured power falls between 0 and $P$ for an assumed $P_s$.  

\begin{equation}
f_{n}(P;P_{s}) = \int_{0}^{P} p_{n}(x;P_{s})\,dx
\label{eq:int}
\end{equation}

We calculated upper limits on intrinsic pulsed fraction by first using Equation~\ref{eq:int} together with the maximum measured power $P_{\rm max}=0.5\,Z_{1}^{2}$(max)
found in a period search to compute the $99\%$ confidence upper limit on intrinsic power $P_s$, i.e., the value of $P_s$ such that $f_n(P_{\rm max};P_s)=0.99$.
We then computed the corresponding upper limit on pulsed fraction, $f_{p}^{\rm max}$, using the \citet{pav99} relation between pulsed fraction, power, and the total number of counts $N$ for the special case of sinusoidal signals:
\begin{equation}
f_{p}^{\rm max} = 2(1 + N_{b}/N_{s}) \sqrt{P_s/N},
\label{eq:z2_to_pf}
\end{equation}
where $N_{b}/N_{s}$ is the ratio of background counts to source counts in the source extraction aperture, and $N_s+N_b=N$.

For an observation of length $T$ and time resolution $\delta t$, the smallest detectable frequency is $f_{\rm min} \equiv 1/T$, and the largest detectable frequency is the Nyquist frequency $f_{\rm max} \equiv 1/ 2 \delta t$. 
The number of independent Fourier frequencies is equal to $T (f_{\rm max} - f_{\rm min})$.
In practice, a periodicity search may miss a weak signal in the $Z_{1}^{2}$ statistic when the peak occurs between two independent Fourier frequencies. To ensure that we do not miss a signal in between two independent Fourier frequencies, we oversampled the independent frequencies by a factor of at least 5 in all of our searches.
This oversampling, while necessary, would complicate our calculation of the upper limits on the CCO pulsed fractions if we used $Z_{1}^{2}$(max) from the oversampled searches.
To remedy this, we use only $Z_{1}^{2}$(max) from the independent Fourier frequencies to compute pulsed fraction upper limits.
When searching for pulsations in multiple observations, we added the power spectra, and used the $Z_{1}^{2}$(max) from the total power spectrum to calculate the upper limit on the pulsed fraction.

Table~\ref{tbl:pf_upper_limits} lists the updated upper limits on pulsed fraction from this timing analysis.
We conclude from this that the CCO in Kes~79 is an outlier, i.e.,that none of the other confirmed CCOs have similarly large pulsed fractions of $64\%$.
The only caveat concerns CXOU J160103.1$-$513353; its pulsed fraction upper limit is $<50 \%$ for all plausible periods down to 12 ms, and $<25 \%$ for periods greater than 146~ms (the Nyquist limit of the \xmm \ pn detector in full window mode).

The CCO in Cas A was observed for 487~ks over a period of 11 days in March 2009 with the Chandra HRC-S. 
Because of the close spacing of these observations, they can be searched coherently. 
\cite{hal10a} reported on a search of the 433~ks of data that were publicly available at that time, and calculated as $12 \%$ upper limit on pulsed fraction.  Here we analyze the full set of observations.
We searched the ($f,\dot f$) parameter space over the region  $f \le 200$~Hz for $\dot f \le 5 \times 10^{-13}$~Hz~s$^{-1}$, $f \le 100$~Hz for $\dot f \le 3 \times 10^{-12}$~Hz~s$^{-1}$, and $f \le 10$~Hz for $\dot f \le 1 \times 10^{-10}$~Hz~s$^{-1}$, oversampling by a factor of $\approx 3$ in each parameter.
With the addition of this new data, we find a slightly lower $11 \%$  upper limit on pulsed fraction.

The four non-pulsing CCOs whose spectra are consistent with UTCA models (see Figures~\ref{fig:g330_g350_g15_spec} and \ref{fig:cas_a_spec}), also happen to have the largest upper limits on pulsed fraction listed in Table~\ref{tbl:pf_upper_limits}.
So, to the extent that the spectra of any CCOs are consistent with a UTCA, their upper limits on pulsed fraction add little support to the argument.
The best case for a UTCA can be made for the CCO in Cas A.  However, its $11\%$  upper limit on pulsed fraction is still not lower than the measured $11\%$ and $9\%$ amplitude pulsations of \pupa \ and \twelveohseven.

The CCOs in Vela Jr. and \gthreefortysevensnr \ have two of the three lowest pulsed fraction upper limits ($5\%$ and $3\%$, respectively).
We showed in Section \ref{section:vela_jr_g347} that their distances are too close to have uniform-temperature surfaces accounting for their spectra.
The third CCO with a very low pulsed fraction upper limit is the CCO in \gthreefiftythreesnr, with an upper limit of $5\%$.
In Section \ref{section:g353_spec}, we demonstrated its X-ray spectrum is inconsistent with a UTCA for any reasonable distance and NS radius.

Therefore, the three CCOs with the smallest upper limits on pulsed fraction evidently do not have UTCAs.
Pulsed fraction upper limits of $5\%$ or even $3\%$ are therefore not proof of a uniform-temperature surface, let alone the larger upper limits for the other CCOs with published UTCA fits.  The CCOs in Vela Jr. and \gthreefortysevensnr, plus the three CCOs with measured X-ray pulses, comprise five out the eight well-established CCOs that are better understood as having localized thermally emitting surface regions, even while four of these five have pulsed fractions $\leq11\%$.
Since uniform-temperature surfaces are disfavored for these CCOs, in the following section we consider what alternative conclusions can be drawn.

\section{Constraints on CCO hot-spots}
\label{section:spot_modeling}

We have shown that some CCOs without detected pulsations are still expected to have small hot-spots based on their spectra.
Their pulse modulation is simply below current detection limits.
A small pulse modulation indicates some combination of 1) a small angle $\psi$ between the rotation axis and the observer's line of sight, and 2) a small inclination angle $\xi$ between the rotation axis and the hot-spot pole.
Viewing angle $\psi$ is sinusoidally distributed: $D(\psi) = \sin(\psi)$. 
The reason is that the distribution of angles obtained by sampling the angles between two vectors pointing in random directions in 3D space is a sine distribution.
In this case the two vectors are the NS spin axis and our line of sight to the NS, and there is no known selection effect that would bias the viewing angle.

The calculations by \cite{sul17} and \cite{dor18} assume that $\xi$ is also sinusoidally distributed: $D(\xi) = \sin(\xi)$, with hot-spots randomly distributed in 3D in space, therefore more likely located near the rotational equator where they can produce stronger X-ray pulses.  
This amounts to an assumption that the strong crustal magnetic fields producing CCO hot-spots are physically uncorrelated with the NS rotation axis.  
Their statistical conclusion that some CCOs must have UTCAs then depends critically on this physical assumption.

\begin{deluxetable*}{lccccccccccc}
\tablecaption{Hot-spot Angular Radii $\beta$ Implied by Blackbody Models and/or Pulse Modeling}
\tablehead{
\colhead{SNR}  & \colhead{$N_{\rm H}$} &  \colhead{$T_{1}$} &  \colhead{$L_{1}$} & \colhead{$T_{2}$} &  \colhead{$L_{2}$} & \colhead{$D$ (fixed)} & \colhead{$\beta_{1}$} & \colhead{$\beta_{2}$}   & \colhead{$\chi^{2}_{\nu}$ (d.o.f.)} & \colhead{$H_{0}$}\\
\colhead{}  & \colhead{($10^{22}$ cm$^{-2}$)} &  \colhead{(keV)} & \colhead{($10^{33}$ erg s$^{-1}$)} & \colhead{(keV)} & \colhead{($10^{33}$ erg s$^{1}$)}  & \colhead{(kpc)} & \colhead{($^{\circ}$)} & \colhead{($^{\circ}$)}  & \colhead{} & \colhead{} }
\startdata
Puppis~A\tablenotemark{a}  & 0.58$_{-0.02}^{+0.01}$  & 0.222$_{-0.018}^{+0.019}$ & 0.95$_{-0.03}^{+0.04}$ & 0.411$_{-0.011}^{+0.011}$ & 1.01$_{-0.01}^{+0.01}$ &  1.3 & 15.23 & 3.93 & 1.01 (284) & 0.442 \\
Vela Jr. &  0.334$_{-0.012}^{+0.012}$ &  0.383$_{-0.004}^{+0.004}$  &  0.141$_{-0.002}^{+0.002}$   & ... & ... &  0.75  &  0.76  & ...   &  1.172 (56)  &  0.178 \\ 
PKS 1209$-$51/52 & 0.135$_{-0.021}^{+0.021}$ &  0.147$_{-0.006}^{+0.006}$  &  2.789$_{-0.73}^{+0.73}$   &  0.35$_{-0.066}^{+0.066}$  &  0.843$_{-0.173}^{+0.173}$  &  2.1  &  23.45  &  2.23  &  1.251 (43)  &  0.126 \\
G330.2$+$1.0 &  2.839$_{-0.296}^{+0.296}$ &  0.407$_{-0.03}^{+0.03}$  &  1.934$_{-0.421}^{+0.421}$   & ... & ... &  4.9  &  2.5  & ...   &  1.156 (50)  &  0.209 \\ 
G347.3$-$0.5 &  0.474$_{-0.026}^{+0.026}$ &  0.251$_{-0.032}^{+0.032}$  & 0.529$_{-0.074}^{+0.074}$    &  0.441$_{-0.024}^{+0.024}$  &  0.783$_{-0.138}^{+0.138}$  &  0.13  &  0.34  &  0.14  &  1.179 (56)  &  0.17 \\
G350.1$-$0.3 &  3.113$_{-0.508}^{+0.508}$ &  0.472$_{-0.064}^{+0.064}$  &  4.145$_{-1.243}^{+1.243}$   & ... & ... &  4.5  &  2.72  & ...   &  0.843 (24)  &  0.683 \\ 
Kes~79\tablenotemark{a}  & 1.52 (fixed)  &  0.223$_{-0.042}^{+0.042}$  & 6.56 &   0.516$_{-0.042}^{+0.042}$ & 12.1  &  7.1  &  28.6  &  7.2  &  0.9 (31)  &  0.626    \\
Cas~A &  1.372$_{-0.06}^{+0.06}$ &  0.372$_{-0.009}^{+0.009}$  &  3.101$_{-0.204}^{+0.204}$   & ... & ... &  3.33  &  3.79  & ...   &  0.968 (135)  &  0.588 \\ 
\hline
G353.6$-$0.7\tablenotemark{b}  & 1.492$_{-0.044}^{+0.042}$  &  0.487$_{-0.005}^{+0.005}$  &  7.584$_{-0.131}^{+0.131}$   &  ...  &  ...  &  3.2  &  3.92  &  ...  &  1.155 (74)  &  0.171  \\
G15.9+0.2 &  4.516$_{-1.077}^{+1.077}$ &  0.342$_{-0.068}^{+0.068}$  &  3.778$_{-3.412}^{+3.412}$   & ... & ... &  8.5  &  4.93  & ...   &  0.547 (17)  &  0.93 \\ 
\enddata
\tablecomments{
The NS mass and radius are held fixed at 1.4~$\msun$ and 12~km when calculating the hot-spot angular radii $\beta_{1}$ and $\beta_{2}$.
Hot-spot angular radii $\beta_{1}$ and $\beta_{2}$ are calculated assuming there are two identical antipodal emitting spots, consisting of a hot spherical cap with temperature $T_{2}$, and an annular region with a lower temperature $T_{1}$. 
Distances are held fixed, and chosen for consistency with the independently measured values listed in Table \ref{tbl:cco_distances}.}
\tablenotetext{a}{
Parameters for Puppis~A and Kes~79 are taken from \cite{alf22} and \cite{bog14}, respectively, with the values for Kes~79 corresponding to a magnetic hydrogen atmosphere model.}
\tablenotetext{b}{The X-ray spectrum of the CCO in  G353.6$-$0.7 is well described by a two blackbody model, with the cooler component likely originating from the whole NS surface. The cooler blackbody temperature is not well constrained by the data, and here we have held it fixed at 0.08 keV.}
\label{tbl:blackbody_results}
\end{deluxetable*}

Next, we will make a rough estimate of the distribution $D(\xi)$ that would be consistent with the upper limits on pulsed fraction calculated in Section~\ref{section:timing}, under the assumption that all CCOs have small antipodal hot-spots.
First we will use the general relativistic emission model described in \cite{alf22} to show that, if a given CCO does have hot-spots, then the spots are small enough that their effect on the light curve is as if they are effectively point-like.
Then we will calculate the observed pulsed fractions from an emission model with two small, identical, antipodal hot-spots, as a function of the geometric angles $\psi$ and $\xi$.
Finally, we will calculate the probability of detecting X-ray pulsations from a randomly selected CCO, for a given assumed $D(\xi)$ and an observed upper limit on pulsed fraction $f_{p}^{\rm max}$.

Table~\ref{tbl:blackbody_results} lists the spectral parameters resulting from fitting the non-pulsing CCOs to one- or two-temperature blackbody models.
We also list the angular radii $\beta$ of the emitting regions on the NS surface, assuming two identical antipodal spots, for the special case where $\psi = 90^{\circ}$ and $\xi = 0^{\circ}$.
Different hot-spot geometries will have $\beta$ values within a factor of $\sim 2$ of these indicative values.
We model each spot with a hotter circular region with angular radius $\beta_{2}$, and if two blackbodies are required to fit the spectra, a surrounding cooler annular region with angular radius $\beta_{1}$.
For completeness we include the corresponding parameters for the CCOs in Puppis~A and Kes~79, whose energy-dependent pulse profiles have already been successfully modeled \citep{bog14,alf22}.
In the case of Kes~79, it is known that there is a cooler region surrounding a hotter region, but in the case of Puppis~A, the two different temperature regions are actually approximately antipodal.
Angular radii $\beta$ listed in Table~\ref{tbl:blackbody_results} should be considered a rough estimate of the emitting region sizes.

For all of the non-pulsing CCOs in Table~\ref{tbl:blackbody_results} $\beta \lesssim 10^{\circ}$, so it is a good approximation to treat them as effectively point-like when computing their X-ray pulses.
Also, since the two emitting regions are identical in this model, their different temperatures will not create an energy dependence in the pulse profiles.
For these reasons, assuming isotropic emission, it is sufficient to calculate pulse profiles for just one special case to place constraints on CCO hot-spots.
In the absence of detected X-ray pulsations, our goal is to roughly constrain the sizes and locations of the emitting regions for the CCO population as a whole.

Here we calculate theoretical pulse profiles from a CCO with two identical, single-temperature, antipodal hot-spots with $\beta_{1} = 10^{\circ}$.  
Our emission model predicts the time-dependent X-ray flux, fully including the general relativistic effects of gravitational redshift and light bending. 
We assume a 12 km radius, 1.4 $\msun$ neutron star, with isotropic blackbody emission.
For more details on this emission model, see \cite{alf22}, where it successfully reproduces the energy-dependent pulse profiles of the CCO in Puppis~A.

\begin{figure*}
\centering
\includegraphics[width=1.0\linewidth]{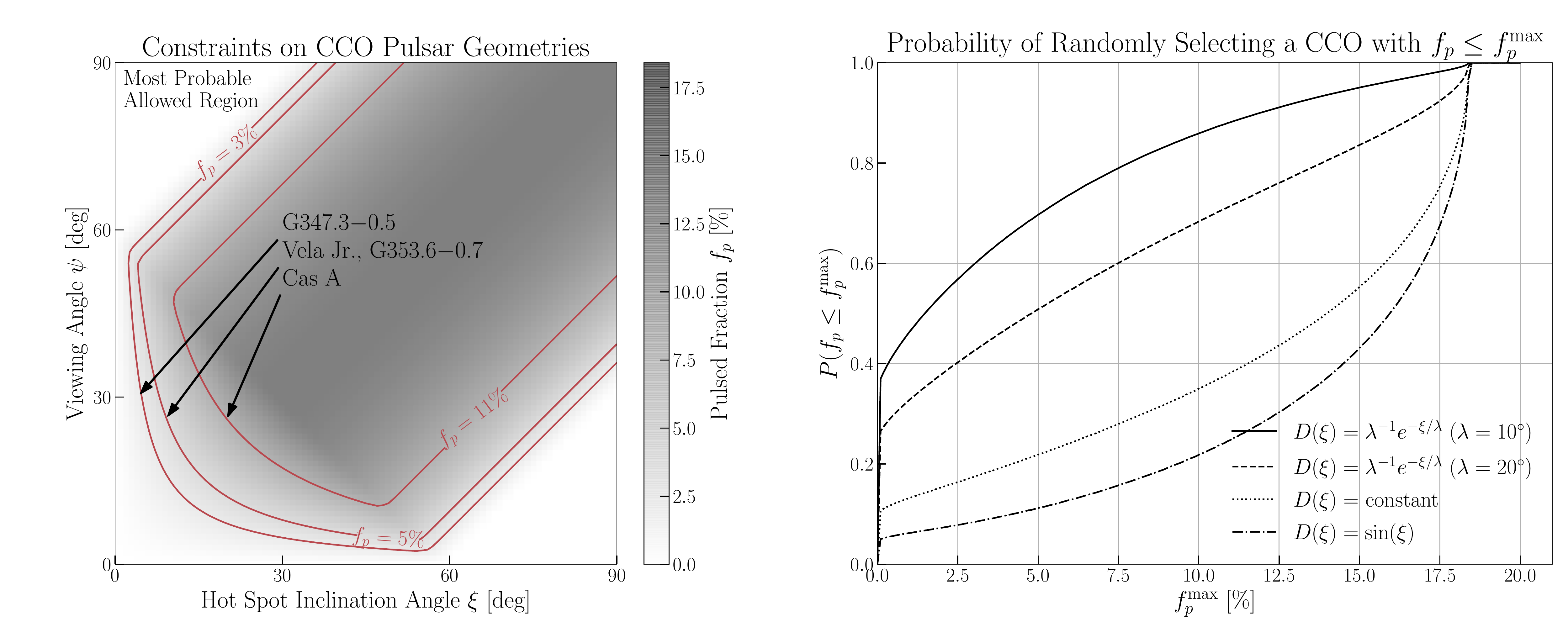}
\caption{Left: Pulsed fractions as a function of the geometric angles $\psi$ and $\xi$, for a CCO emission model with two small identical antipodal hot-spots. 
Pulsed fraction contours indicate the parameter space that is ruled out for several CCOs, because their pulsed fraction upper limits $f_{p}^{\rm max}$ are less than the pulsed fraction $f_{p}$ predicted by the model. 
Right: For various probability distributions $D(\xi)$, we plot the probability that $f_{p} \leq$ $f_{p}^{\rm max}$, for a randomly selected CCO.
}
\label{fig:two_spot_model}
\end{figure*}

Figure \ref{fig:two_spot_model} (left) shows the pulsed fractions calculated from this model as a function of $\xi$ and $\psi$.
We use these pulsed fractions to calculate cumulative distribution functions of pulsed fraction for several assumed statistical distributions $D(\xi)$.
In other words, we calculate the probability that one would expect to have a pulsed fraction less than a given observed upper limit $f_{p}^{\rm max}$, assuming some distribution of $\xi$.
For each $f_{p}^{\rm max}$ we calculate the probability $P(f_{p} \leq f_{p}^{\rm max})$ that a randomly selected CCO has a pulsed fraction less than $f_{p}^{\rm max}$:

\begin{equation}
   P(f_{p} \leq f_{p}^{\rm max}) = \frac{\displaystyle\int_0^{\pi/2}\int_0^{\pi/2}\,S(\psi,\xi)\,\sin \psi\,D(\xi)\, 
  {\rm d}\xi\,{\rm d}\psi}
 {\displaystyle\int_0^{\pi/2}\int_0^{\pi/2}\,\sin \psi\,D(\xi)\,
 {\rm d}\xi\,{\rm d}\psi},
\label{eq:threshold_probability}
\end{equation} 

where $S(\psi,\xi)$ is defined as:

\begin{equation}
S(\psi,\xi) = \left\{\begin{array}{cr} 
1, & f_{p}(\psi,\xi) < f_{p}^{\rm max} \\
0, & f_{p}(\psi,\xi) > f_{p}^{\rm max} \end{array}\right. 
\label{eq:king}
\end{equation}

The right panel in Figure \ref{fig:two_spot_model} shows the cumulative distribution functions calculated from Equation \ref{eq:threshold_probability}, assuming different distributions $D(\xi)$.
Suppose, for example, that the CCO in Vela Jr. has a pulsed fraction just below the $5\%$ upper limit in Table~\ref{tbl:pf_upper_limits}.
Then, a $50\%$ chance of a non-detection is consistent with  $\xi$ being drawn from and exponential distribution with a scale factor $\lambda \sim 20^{\circ}$.
Also, examining the right panel of Figure~\ref{fig:two_spot_model}, we find that if $\xi$ was actually uniformly or sinusoidally distributed then a majority of CCOs would have pulsed fractions greater than $10\%$, in contradiction to what is actually observed.
On the other hand, if $\xi$ is exponentially distributed with a scale factor $\lambda \sim 20^{\circ}$, then is expected that a majority of CCOs will have pulsed fractions $\lesssim10\%$, as is observed.
Six of the ten CCOs that we fit to carbon atmosphere models in Section \ref{section:spec_analysis} have pulsed fractions $\leq 11\%$, which is also roughly consistent an exponential distribution with $\lambda \sim 20^{\circ}$.
We have shown five of these six must have small spots.
Since $\psi$ is sinusoidally distributed, and $D(\xi)$ seems to be consistent with an exponential distribution with  $\lambda \sim 20^{\circ}$, we conclude that the CCOs without detected pulsations are most likely found in the upper left region of Figure~\ref{fig:two_spot_model} (left), with $\xi \lesssim 20^{\circ}$, and  $\psi \gtrsim 60^{\circ}$.

This emission model assumed that the antipodal hot-spots are the same size and temperature, allowing the pulsations to almost perfectly cancel out in some regions of the $\xi$ and $\psi$ parameter space.
If the spots were of unequal sizes, or different temperatures,  then this would slightly increase the pulsed fractions shown in Figure~ \ref{fig:two_spot_model}.
It would not change the conclusion that $\xi$ must be approximately exponentially distributed in order to be consistent with the observed upper limits on pulsed fraction.

\section{Discussion}
\label{section:discusion}

\subsection{Comparison with Previous Studies}
\label{section:disc_prev_studies}

Recently, \cite{wu21} also reported upper limits on the X-ray pulsed fractions of several CCOs.
Their analysis differs from ours in three ways.
First, \cite{wu21} did not account for the known background count rates, i.e., they did not include the necessary factor $(1 + N_{b}/N_{s})$ of our Equation \ref{eq:z2_to_pf}.
This is important for the CCOs in \velajrsnr, \gthreethirtysnr, \gthreefiftysnr, and \gthreefiftythreesnr, which have $N_b/N\geq 0.1$, and especially in the case of the CCO in \gthreethirtysnr, which has $N_b/N$ as high as 0.47.
Second, they only searched for pulsations in the full energy range of each CCO X-ray spectrum.
It is useful also to search for pulsations in multiple energy bands, in order to not miss weak signals that feature energy-dependent phase shifts, such as those produced by the CCO in Puppis~A.
Third, they did not account for the effects of noise in the power spectrum by using the appropriate probability distribution.
In searching for weak signals, noise is always a significant contribution to the total measured power.
 \cite{wu21} reported a 18$\%$ upper limit ($3 \sigma$) for the CCO in \gthreethirtysnr, which is much lower than the 50$\%$ (99$\%$ confidence) upper limit we calculated, likely because they did not take into account the significant ($\sim 40\%$) background rates for this source.
 Their other reported pulsed fraction upper limits are closer to our results, likely because the error introduced by neglecting the  background rates roughly cancelled out the error introduced by neglecting noise power in those cases.

\cite{wu21} performed a similar spectral analysis of CCOs using the {\tt carbatm} model. 
They also found that the distances to \velajrsnr \ and \gthreefortysevensnr \ are inconsistent with emission from the whole surface of a NS with a reasonable mass and radius.
\cite{wu21} choose not to analyze the longest observation of the CCO in \gthreefiftythreesnr \ (ObsID 0722190201), writing that it was hard to select a background region.
We found that the background rate was not too large during this observation ($\sim 4\%$ for the spectral data analysis), and that this high quality spectral data demonstrates that the {\tt carbatm} model is not consistent with the spectrum of this CCO.

\cite{sul17}, \cite{dor18}, and \cite{wu21} performed statistical calculations involving the distribution of hot-spots similar to our calculations in Section \ref{section:spot_modeling}.
Their calculations assume that the hot-spot inclination angle $\xi$ is sinusoidally distributed, i.e., random in space.
We demonstrated in Section~\ref{section:spot_modeling} that the upper limits on CCO pulsed fractions, combined with spectral fitting presented in Section~\ref{section:spec_analysis},  invalidate this assumption. 
\cite{sul17} and  \cite{dor18} used a similar model with identical antipodal hot-spots, but their model features anisotropic emission, as expected from a non-magnetic hydrogen atmosphere.
This will tend to produce slightly larger pulsed fractions than our isotropic emission model. 

\cite{dor18} and \cite{wu21} calculated theoretical pulsed fractions for viewing angles $0^{\circ} \leq \psi \leq 90^{\circ}$, not the full $180^{\circ}$ range.
Their models features two antipodal spots with different temperatures and sizes, but since they did not consider the full range of $\psi$, their results are biased by the arbitrary choice of whether the hotter spot is the ``near'' one or
the ``far'' one.
Our modeling, and the modeling of \cite{sul17}, uses identical antipodal emitting regions, so the results are symmetric with respect to the interchange of the two hot-spots, and it sufficient to calculate pulsed fractions in the restricted range $0^{\circ} \leq \psi \leq 90^{\circ}$.

In the absence of detected X-ray pulsations, it is impossible to know what type of emission model is more realistic for a given CCO.
If the spectrum can be described by a two-blackbody model, the two temperatures could still be coming from adjacent regions of the NS surface, as \cite{bog14} demonstrated is the case for the CCO in \kesseventyninesnr.
Or, the two temperatures could actually correspond to two disconnected regions on the NS surface, as \cite{alf22} demonstrated is the case for the CCO in \pupasnr.
The emission from the CCO in \pupasnr \ is consistent with an isotropic emission pattern, while the emission from the CCO in \kesseventyninesnr \ is consistent with a very anisotropic emission pattern.
Nevertheless, our modelled pulsed fractions on the left of Figure \ref{fig:two_spot_model} are still similar to the model results of \cite{sul17} and \cite{dor18}, which assume anisotropic emission.
It is the assumption that $\xi$ is randomly distributed in space, or not, that accounts for our different conclusions about CCOs having uniform-temperature surfaces.  
Since our spectral fits presented in Section~\ref{section:spec_analysis} show that the surface temperatures of \velajr \ and \gthreefortyseven \ cannot be uniform, we conclude that $\xi$ is likely not random.

\cite{dor18} actually contemplated the possibility that some process aligns the magnetic field of the neutron star with the rotation axis, but
wondered why the same process would not affect the
pulsating CCO population.  
We have evidence that there is such an effect, and that it also aligns the hot-spots seen on the pulsating CCOs.
\cite{alf22} found that the primary hot-spot on \pupa \ is likely very close to its rotation axis, with a most probable value of $\xi \approx 6^{\circ}$.
\cite{bog14} found that an emission model with a hot-spot $<10^{\circ}$ from the rotation axis can reproduce the energy-dependent pulse profiles of the CCO in \kesseventyninesnr, while the $64\%$ pulsed fraction is due to highly anisotropic emission.
Anisotropic emission can be produced in strongly magnetized ($\gtrsim 10^{12}$~G) atmospheres (see, e.g. \cite{pav94,zav95}).
A strongly magnetized atmosphere might be expected at the spot where a strong crustal field is channelling heat.
The CCO in \twelveohsevensnr \ has a $9\%$ pulsed fraction, and its pulse profiles show no energy-dependent phase changes. This indicates that the observed emission is likely from just one region on the NS surface, and that this region is likely near the rotational pole.

\cite{pos22} found that a non-magnetic hydrogen atmosphere model fits the Cas A CCO spectra equally as well as the {\tt carbatm} model, with the emitting area of the hydrogen model covering a small fraction of the NS surface.  
They found that while cooling of the atmosphere is implied by the carbon atmosphere model, a hydrogen atmosphere model implies that the size of emitting region(s) on \casa \ are shrinking.

Finally, there are two candidate CCOs whose spectra we have not analyzed in this paper because their X-ray emission is too faint to draw strong conclusions regarding their spectra. 
These candidates are found in the G296.8$-$0.3 and  SN 1987A SNRs \citep{san12,pag20}.

\subsection{Can a CCO avoid gaining one optical depth of H for (at least) hundreds of years?}
\label{section:accretion_rate_estimate}

In order to develop a carbon atmosphere, a CCO must 1) be hot enough for efficient DNB, and also 2) avoid accreting H or He at a rate comparable to the rate that H and He are consumed by DNB. 
A CCO may meet the first criteria since CCOs are very young NSs, but it is not obvious that their positions within SNRs would allow them to meet the critical second criteria of low accretion rates.
The youngest, hottest period of a CCO lifetime occurs when the mean density of the surrounding SNe ejecta is highest.
Also, their very weak pulsar winds, as evidenced by their lack of surrounding pulsar wind nebulae, will not help reduce their accretion rates.

The Bondi-Hoyle rate provides an order-of-magnitude estimate of CCO accretion, given a CCO velocity and the mean density of the surrounding ejecta.
We estimate the mean density in the inner region of a SNR using equations 1 through 3 of \cite{blo01}, derived from the self-similar driven wave solutions of \cite{che82}.
The Cas A SNR in particular is only ${\sim} 350$ years old and therefore its mean ejecta density, which evolves as $t^{-3}$, is still high.
\cite{hob05} found that pulsar birth velocities are well described by Maxwellian distribution with a mean of $400 \pm 40$ km s$^{-1}$, and
\cite{may21} measured CCO proper motions and found that they are consistent with being drawn from this same distribution.
For the CCO in Cas A they measured $v_{\rm proj} = 570 \pm 260$ km s$^{-1}$, consistent with a previous $v_{\rm proj} \approx 350 $ km s$^{-1}$ measurement based on a determination of the kinematic center of the SNR \citep{fes06,tho01}.
Observations and modeling of Cas A indicate that the explosion energy $E_{\rm SN} \approx 2 \times 10^{51}$ erg and the ejecta mass $M_{\rm ej} \approx 3 \msun$ \citep{orl16,vin96,wil03,lam20}.
Assuming the NS velocity $v_{\rm{NS}}$ is much greater than the sound speed, the Bondi-Hoyle accretion rate for a $1.4 \ \msun$ NS is:

$$\dot{M} = 1.2 \times 10^{14} \left(\frac{E_{\rm{SN}}}{2 \times 10^{51} \ \rm{erg} }\right)^{-3/2} \left(\frac{M_{\rm{ej}}}{3\,\msun }\right)^{5/2} \left(\frac{t}{350 \ \rm{yr}}\right)^{-3}  \left(\frac{v_{\rm{NS}}}{500 \ \rm{km} \ \rm{s}^{-1}}\right)^{-3}  \ \ \rm{g} \ \rm{yr}^{-1}.$$

Therefore, in order to produce and maintain a carbon atmosphere, the ejecta density along the Cas A CCO trajectory would have to be approximately three orders of magnitude less than the mean SNR density.
While the accreted ejecta is  unlikely to be pure H, \cite{bil92} found that a NS accreting CNO elements will produce significant atmospheric H through spallation.

\subsection{Effects of Magnetic Field Structure and Evolution on Surface Thermal Emission}
\label{ssec:disc_physics}

Localized hot-spots on NSs are naturally explained by anisotropic heat conduction due to strong crustal magnetic fields (see, e.g., \citealt{gre83}).
Thermal conductivity is higher parallel to the $B$-field; radially directed crustal fields will be thermally connected to the NS interior, and crustal $B$-fields parallel to the NS surface will act as insulators.

It in unclear if the hot-spots on CCOs correspond to the location of the dipole field axis, or to other field components.
In either case, spin-down measured dipole $B$-fields for CCOs are smaller than those of normal radio pulsars.
COOs may have intrinsically weak dipole fields, or they may have been born with dipole fields comparable to normal radio pulsars, with their fields promptly ``buried'' by hypercritical accretion following the supernova explosion.
Here, we will consider the intrinsically weak dipole field, and buried dipole field scenarios separately.

In the first scenario, CCOs could simply be born with weak global dipole fields, and strong, localized crustal magnetic fields, preferentially formed close to their rotation axes.
Dynamos are likely necessary to create magnetar strength fields (see, e.g., \cite{spr08}), and correlations between rotation and magnetic fields are expected both during pre-collapse stellar evolution \citep{heg05}, and in proto-neutron stars \citep{nag20}.
An advantage of this scenario is that the same physical processes can create both the weak dipole field and the stronger localized components, and correlations with NS rotation rate and rotation axis could arise naturally in a dynamo mechanism.
Magnetic field amplification in NS births is an active area of research, and the details of such a mechanism would need to be worked out in detail.

There are two primary challenges for this intrinsically weak dipole field scenario.
First, \cite{kas10} noted that the CCOs in \kesseventyninesnr\ and \pupasnr\ are in an underpopulated region of the $P-\dot{P}$ diagram.
If CCOs are common, this region of the region of the $P-\dot{P}$ diagram, should be dense with their descendants.
\cite{kas10} noted that selection effects cannot explain the under-density of radio pulsars in this region of the $P-\dot{P}$ parameter space.
A plausible solution is that NS radio luminosity may be dependent on spin-down power \citep{fau06}, creating a selection bias against discovering NSs in region of the $P-\dot{P}$ diagram occupied by CCOs. 

The second missing element of the intrinsically weak dipole field scenario is a mechanism that produces weaker CCO dipole fields than the dipole fields of normal radio pulsars.
\cite{fau06} found that birth period distribution of normal radio pulsars has a mean of $\sim 300$~ms and a dispersion of $\sigma\sim 150$~ms.
So the birth spin periods of CCOs may be comparable to normal radio pulsars, and this parameter alone would not explain their different spin-down measured dipole fields.
The difference in dipole field strength would have to depend on the progenitor's seed field, or some other unknown parameter. 

Now, we consider the second scenario, where CCOs are born with  $B$-fields comparable to normal radio pulsars, and then these fields are buried by hypercritical accretion following the supernova explosion.
In this scenario, the under-density of radio pulsars in the $P-\dot{P}$ diagram near CCOs is naturally explained: the eventual re-emergence of the buried field increases $\dot{P}$ and the CCO moves up the diagram toward the normal radio pulsars.

If CCOs are produced by $B$-field burial, then their hot-spots likely correspond to the poles of the buried dipole field.
Here we discuss what mechanism could be aligning their dipole fields with their rotation axes, to produce the correlation observed in CCOs.
It has long been known that electromagnetic torques can decrease the magnetic inclination angle of a NS, with calculations by \cite{mic70} showing that these torques can align the rotational and magnetic axes, initially separated by an angle $\xi_0$ on a NS in vacuum, on an exponential timescale $\tau^{\rm vac}_{\rm align}$ comparable to the spin-down timescale $\tau_0$:

\begin{equation}
\tau^{\rm vac}_{\rm align}=1.5\tau_0\cos^{-2}\xi_0,
\end{equation}
where
\begin{equation}
\tau_0 = \frac{I c^3}{\mu^2 \Omega^2_0} \approx 10^8 \left(\frac{B_0}{10^{11}\, {\rm
    G}}\right)^{-2} \left(\frac{P_0}{100\, {\rm ms}}\right)^{2}~{\rm yr},
\label{eq:tau0}
\end{equation}
and the sine of the inclination angle decreases exponentially

\begin{equation}
\sin\xi = \sin \xi_0 \exp \left(-{t}/{\tau^{\rm vac}_{\rm align}}\right).
\label{eq:sin_xi}
\end{equation}

More recently, \cite{phi14} calculated the change of inclination angle using MHD simulations that include magnetospheric torques, and found that the inclusion of magnetospheric effects \emph{decreases} the rate of change of inclination angle from an exponential to a power law. 
The calculations by \cite{mic70} and \cite{phi14} show that electromagnetic and magnetospheric torques act too slowly to significantly change the inclination angles of CCOs from their birth values, since CCOs are only hundreds or thousands of years old, as determined by their associations with SNRs.

A faster alignment mechanism was presented by \cite{dal17}, who studied the coupled interior viscosity and magnetic field evolution of a NS, and found that it could produce small inclination angles on $\sim$300 year timescales.  
This mechanism is therefore fast enough to produce small inclination angles on CCOs.
Whether the mechanism increases or decreases the initial inclination angle
depends on the magnetically-induced ellipticity, mass, radius, and initial spin period, so it may be the case that the values of these parameters applicable to CCOs tend to produce small inclination angles.
For a 1.4 $\msun$, 12~km radius CCO, with a $\sim100$~ms spin period similar to the CCOs in \kesseventyninesnr \ and \pupasnr, this mechanism will produce a small inclination angle if the magnetically-induced ellipticity $\epsilon_B\gtrsim$~a few $\times 10^{-7}$. 
For a longer spin period, such as the 424~ms spin period of the CCO in \twelveohsevensnr, a small inclination angle would be produced if the magnetically-induced ellipticity $\epsilon_B\gtrsim$~a few $\times 10^{-8}$. 
A large buried field could then explain both the existence of hot-spots on CCOs, and their small inclination angles.

\subsection{Directions for Future Work}

A better understanding of the physical properties of CCOs will require more detections  of their rotational periods and period derivatives.
These measurements can be made with either longer X-ray timing observations, or a new instrument.  There is a specific need for a new instrument that could mask out the very bright SNR emission surrounding the Cas~A CCO, while retaining the timing, spatial resolution, and throughput of (at least) \chandra.
If more rotation periods are detected, then energy-dependent pulse profile modeling can further constrain the surface emission geometries of CCOs.

CCOs are perhaps the least understood class of young NSs.
This poses a theoretical challenge, since CCOs represent a significant fraction of NS births.
Somehow, a significant fraction of young NSs end up with low spin-down measured dipole fields, and larger, crustal fields preferentially located near their rotational poles.
This correlation requires further investigation, and a theoretical explanation.

\section{Summary}
We have shown that the CCOs in G347.3$-$0.5, Vela Jr., and G353.6$-$0.7 must have non-uniform temperature surfaces with one or more small hot spots, as opposed to UTCAs. 
This is because G347.3$-$0.5 and Vela Jr. are located at distances too close for a uniform-temperature {\tt carbatm} model to give a reasonable NS radius, and the X-ray spectrum of G353.6$-$0.7 is inconsistent with the {\tt carbatm} model.
Since the three pulsed CCOs in Puppis A, Kes 79, and PKS 1209$-$51/52 were previously known to have multi-temperature surfaces, at least six of the dozen known CCOs must have multi-temperature surfaces.
We demonstrated that the X-ray spectra of the three pulsed CCOs can be fitted with UTCA models of appropriate NS area, despite having manifestly non-uniform surface temperatures.
We then calculated new upper limits on CCO X-ray pulse modulation using the best available data on the unpulsed CCOs.
Taken together, our spectral and timing analyses indicate that there is no definitive evidence that any of the remaining unpulsed CCOs has a UTCA.
To account for the possibility that all CCOs have localized, hot regions on the NS surface, we developed a model where these hot-spots are located preferentially near the rotational poles, and are consistent with an exponential distribution with a scale factor $\lambda \sim 20^{\circ}$.
Strong crustal magnetic fields are required to produce CCO hots spots, and the correlation of these magnetic fields with the NS rotation axis requires further investigation.

\begin{acknowledgements}
We thank the anonymous referee for useful comments.
JA would like to thank Malvin Ruderman for helpful discussions regarding magnetic effects in NS atmospheres.
Support for this work was provided by NASA through XMM grant 80NSSC21K0819 and Chandra Award SAO GO0-21059X issued by the Chandra X-ray Observatory Center, which is operated by the Smithsonian Astrophysical Observatory for and on behalf of NASA under contract NAS8-03060.
This investigation is based on observations obtained with XMM-Newton, an ESA science mission with instruments and
contributions directly funded by ESA Member States and NASA.
\end{acknowledgements}

\facilities{XMM, CXO}

\software{{\tt astropy} \citep{ast13, ast18}, {\tt matplotlib} \citep{hun07}, {\tt xspec} \citep{arn96}, {\tt numpy} \citep{har20}, {\tt scipy} \citep{vir20}} 


\begin{thebibliography}{}

\bibitem[Alarie et al.(2014)]{ala14} Alarie, A., Bilodeau, A., \& Drissen, L.\ 2014, \mnras, 441, 2996. doi:10.1093/mnras/stu774
\bibitem[Alcock \& Illarionov(1980)]{alc80} Alcock, C. \& Illarionov, A.\ 1980, \apj, 235, 534. doi:10.1086/157656
\bibitem[Alford et al.(2022)]{alf22} Alford, J.~A.~J., Gotthelf, E.~V., Perna, R., et al.\ 2022, \apj, 927, 233. doi:10.3847/1538-4357/ac4d9a
\bibitem[Allen et al.(2015)]{all15} Allen, G.~E., Chow, K., DeLaney, T., et al.\ 2015, \apj, 798, 82. doi:10.1088/0004-637X/798/2/82
\bibitem[Arnaud(1996)]{arn96} Arnaud, K.~A.\ 1996, Astronomical Data Analysis Software and Systems V, 101, 17
\bibitem[Astropy Collaboration et al.(2013)]{ast13} Astropy Collaboration, Robitaille, T.~P., Tollerud, E.~J., et al.\ 2013, \aap, 558, A33. doi:10.1051/0004-6361/201322068
\bibitem[Astropy Collaboration et al.(2018)]{ast18} Astropy Collaboration, Price-Whelan, A.~M., Sip{\H{o}}cz, B.~M., et al.\ 2018, \aj, 156, 123. doi:10.3847/1538-3881/aabc4f
\bibitem[Blondin et al.(2001)]{blo01} Blondin, J.~M., Chevalier, R.~A., \& Frierson, D.~M.\ 2001, \apj, 563, 806. doi:10.1086/324042
\bibitem[Bignami et al.(2003)]{big03} Bignami, G.~F., Caraveo, P.~A., De Luca, A., et al.\ 2003, \nat, 423, 725. doi:10.1038/nature01703
\bibitem[Bildsten et al.(1992)]{bil92} Bildsten, L., Salpeter, E.~E., \& Wasserman, I.\ 1992, \apj, 384, 143. doi:10.1086/170860
\bibitem[Blondin(1986)]{blo86} Blondin, J.~M.\ 1986, \apj, 308, 755. doi:10.1086/164548
\bibitem[Bogdanov(2014)]{bog14} Bogdanov, S.\ 2014, \apj, 790, 94. doi:10.1088/0004-637X/790/2/94
\bibitem[Braun et al.(2019)]{bra19} Braun, C., Safi-Harb, S., \& Fryer, C.~L.\ 2019, \mnras, 489, 4444. doi:10.1093/mnras/stz2437
\bibitem[Buccheri et al.(1983)]{buc83} Buccheri, R., Bennett, K., Bignami, G.~F., et al.\ 1983, \aap, 128, 245
\bibitem[Cassam-Chena{\"\i} et al.(2004)]{cas04} Cassam-Chena{\"\i}, G., Decourchelle, A., Ballet, J., et al.\ 2004, \aap, 427, 199. doi:10.1051/0004-6361:20041154
\bibitem[Chang et al.(2004)]{cha041} Chang, P., Arras, P., \& Bildsten, L.\ 2004, \apjl, 616, L147. doi:10.1086/426789
\bibitem[Chang et al.(2010)]{cha10} Chang, P., Bildsten, L., \& Arras, P.\ 2010, \apj, 723, 719. doi:10.1088/0004-637X/723/1/719
\bibitem[Chang \& Bildsten(2004)]{cha042} Chang, P. \& Bildsten, L.\ 2004, \apj, 605, 830. doi:10.1086/382271
\bibitem[Chang \& Bildsten(2003)]{cha03} Chang, P. \& Bildsten, L.\ 2003, \apj, 585, 464. doi:10.1086/345551
\bibitem[Chevalier(1982)]{che82} Chevalier, R.~A.\ 1982, \apj, 258, 790. doi:10.1086/160126
\bibitem[Dall'Osso \& Perna(2017)]{dal17} Dall'Osso, S. \& Perna, R.\ 2017, \mnras, 472, 2142. doi:10.1093/mnras/stx2097
\bibitem[de Luca(2008)]{del08} de Luca, A.\ 2008, 40 Years of Pulsars: Millisecond Pulsars, Magnetars and More, 983, 311. doi:10.1063/1.2900173
\bibitem[de Luca et al.(2012)]{del12} de Luca, A., Salvetti, D., Sartori, A., et al.\ 2012, \mnras, 421, L72. doi:10.1111/j.1745-3933.2011.01209.x
\bibitem[De Luca(2017)]{del17} De Luca, A.\ 2017, Journal of Physics Conference Series, 932, 012006. doi:10.1088/1742-6596/932/1/012006
\bibitem[Doroshenko et al.(2018)]{dor18} Doroshenko, V., Suleimanov, V., \& Santangelo, A.\ 2018, \aap, 618, A76. doi:10.1051/0004-6361/201833271
\bibitem[Doroshenko et al.(2022)]{Doroshenko2022} Doroshenko, V., Suleimanov, V., P{\"u}hlhofer, G., et al.\ 2022, Nature Astronomy, 6, 1444. doi:10.1038/s41550-022-01800-1
\bibitem[Faucher-Gigu{\`e}re \& Kaspi(2006)]{fau06} Faucher-Gigu{\`e}re, C.-A. \& Kaspi, V.~M.\ 2006, \apj, 643, 332. doi:10.1086/501516
\bibitem[Fesen et al.(2006)]{fes06} Fesen, R.~A., Hammell, M.~C., Morse, J., et al.\ 2006, \apj, 645, 283. doi:10.1086/504254
\bibitem[Fraija \& Bernal(2015)]{fra15} Fraija, N. \& Bernal, C.~G.\ 2015, \mnras, 451, 455. doi:10.1093/mnras/stv1015
\bibitem[Gaensler et al.(2008)]{gae08} Gaensler, B.~M., Tanna, A., Slane, P.~O., et al.\ 2008, \apjl, 680, L37. doi:10.1086/589650
\bibitem[Geppert et al.(2004)]{gep04} Geppert, U., K{\"u}ker, M., \& Page, D.\ 2004, \aap, 426, 267. doi:10.1051/0004-6361:20040455
\bibitem[Giacani et al.(2009)]{gia09} Giacani, E., Smith, M.~J.~S., Dubner, G., et al.\ 2009, \aap, 507, 841. doi:10.1051/0004-6361/200912253
\bibitem[Giacani et al.(2000)]{gia00} Giacani, E.~B., Dubner, G.~M., Green, A.~J., et al.\ 2000, \aj, 119, 281. doi:10.1086/301173
\bibitem[Goldreich(1970)]{gol70} Goldreich, P.\ 1970, \apjl, 160, L11. doi:10.1086/180513
\bibitem[Gotthelf et al.(2005)]{got05} Gotthelf, E.~V., Halpern, J.~P., \& Seward, F.~D.\ 2005, \apj, 627, 390. doi:10.1086/430300
\bibitem[Gotthelf \& Halpern(2009)]{got09} Gotthelf, E.~V. \& Halpern, J.~P.\ 2009, \apjl, 695, L35. doi:10.1088/0004-637X/695/1/L35
\bibitem[Gotthelf et al.(2010)]{got10} Gotthelf, E.~V., Perna, R., \& Halpern, J.~P.\ 2010, \apj, 724, 1316. doi:10.1088/0004-637X/724/2/1316
\bibitem[Gotthelf et al.(2013)]{got13} Gotthelf, E.~V., Halpern, J.~P., \& Alford, J.\ 2013, \apj, 765, 58. doi:10.1088/0004-637X/765/1/58
\bibitem[Greenstein \& Hartke(1983)]{gre83} Greenstein, G. \& Hartke, G.~J.\ 1983, \apj, 271, 283. doi:10.1086/161195
\bibitem[Groth(1975)]{gro75} Groth, E.~J.\ 1975, \apjs, 29, 285. doi:10.1086/190343
\bibitem[Halpern \& Gotthelf(2010)]{hal10a} Halpern, J.~P. \& Gotthelf, E.~V.\ 2010, \apj, 709, 436. doi:10.1088/0004-637X/709/1/436
\bibitem[Halpern \& Gotthelf(2010)]{hal10b} Halpern, J.~P. \& Gotthelf, E.~V.\ 2010, \apj, 710, 941. doi:10.1088/0004-637X/710/2/941
\bibitem[Harris et al.(2020)]{har20} Harris, C.~R., Millman, K.~J., van der Walt, S.~J., et al.\ 2020, \nat, 585, 357. doi:10.1038/s41586-020-2649-2
\bibitem[Heger et al.(2005)]{heg05} Heger, A., Woosley, S.~E., \& Spruit, H.~C.\ 2005, \apj, 626, 350. doi:10.1086/429868
\bibitem[Heinke \& Ho(2010)]{hei10} Heinke, C.~O. \& Ho, W.~C.~G.\ 2010, \apjl, 719, L167. doi:10.1088/2041-8205/719/2/L167
\bibitem[Ho et al.(2008)]{ho08} Ho, W.~C.~G., Potekhin, A.~Y., \& Chabrier, G.\ 2008, \apjs, 178, 102. doi:10.1086/589238
\bibitem[Ho \& Heinke(2009)]{ho09} Ho, W.~C.~G. \& Heinke, C.~O.\ 2009, \nat, 462, 71. doi:10.1038/nature08525
\bibitem[Ho(2014)]{ho14} Ho, W.~C.~G.\ 2014, Magnetic Fields throughout Stellar Evolution, 302, 435. doi:10.1017/S1743921314002683
\bibitem[Ho et al.(2021)]{ho21} Ho, W.~C.~G., Zhao, Y., Heinke, C.~O., et al.\ 2021, \mnras, 506, 5015. doi:10.1093/mnras/stab2081
\bibitem[Hobbs et al.(2005)]{hob05} Hobbs, G., Lorimer, D.~R., Lyne, A.~G., et al.\ 2005, \mnras, 360, 974. doi:10.1111/j.1365-2966.2005.09087.x
\bibitem[Hunter(2007)]{hun07} Hunter, J.~D.\ 2007, Computing in Science and Engineering, 9, 90. doi:10.1109/MCSE.2007.55
\bibitem[Kaspi(2010)]{kas10} Kaspi, V.~M.\ 2010, Proceedings of the National Academy of Science, 107, 7147. doi:10.1073/pnas.1000812107
\bibitem[Klochkov et al.(2016)]{klo16} Klochkov, D., Suleimanov, V., Sasaki, M., et al.\ 2016, \aap, 592, L12. doi:10.1051/0004-6361/201629208
\bibitem[Klochkov et al.(2015)]{klo15} Klochkov, D., Suleimanov, V., P{\"u}hlhofer, G., et al.\ 2015, \aap, 573, A53. doi:10.1051/0004-6361/201424683
\bibitem[Klochkov et al.(2013)]{klo13} Klochkov, D., P{\"u}hlhofer, G., Suleimanov, V., et al.\ 2013, \aap, 556, A41. doi:10.1051/0004-6361/201321740
\bibitem[Laming \& Temim(2020)]{lam20} Laming, J.~M. \& Temim, T.\ 2020, \apj, 904, 115. doi:10.3847/1538-4357/abc1e5
\bibitem[Landstorfer et al.(2022)]{lan22} Landstorfer, A., Doroshenko, V., \& P{\"u}hlhofer, G.\ 2022, \aap, 659, A82. doi:10.1051/0004-6361/202142334
\bibitem[Mayer \& Becker(2021)]{may21} Mayer, M.~G.~F. \& Becker, W.\ 2021, \aap, 651, A40. doi:10.1051/0004-6361/202141119
\bibitem[McClure-Griffiths et al.(2001)]{mcc01} McClure-Griffiths, N.~M., Green, A.~J., Dickey, J.~M., et al.\ 2001, \apj, 551, 394. doi:10.1086/320095
\bibitem[Michel \& Goldwire(1970)]{mic70} Michel, F.~C. \& Goldwire, H.~C.\ 1970, \aplett, 5, 21
\bibitem[Michel(1988)]{1988Natur.333..644M} Michel, F.~C.\ 1988, \nat, 333, 644. doi:10.1038/333644a0
\bibitem[Mori \& Ho(2007)]{mor07} Mori, K. \& Ho, W.~C.~G.\ 2007, \mnras, 377, 905. doi:10.1111/j.1365-2966.2007.11663.x
\bibitem[Nagakura et al.(2020)]{nag20} Nagakura, H., Burrows, A., Radice, D., et al.\ 2020, \mnras, 492, 5764. doi:10.1093/mnras/staa261
\bibitem[Orlando et al.(2016)]{orl16} Orlando, S., Miceli, M., Pumo, M.~L., et al.\ 2016, \apj, 822, 22. doi:10.3847/0004-637X/822/1/22
\bibitem[Page et al.(2020)]{pag20} Page, D., Beznogov, M.~V., Garibay, I., et al.\ 2020, \apj, 898, 125. doi:10.3847/1538-4357/ab93c2
\bibitem[Pavlov et al.(1999)]{pav99} Pavlov, G.~G., Zavlin, V.~E., \& Tr{\"u}mper, J.\ 1999, \apjl, 511, L45. doi:10.1086/311827
\bibitem[Pavlov \& Luna(2009)]{pav09} Pavlov, G.~G. \& Luna, G.~J.~M.\ 2009, \apj, 703, 910. doi:10.1088/0004-637X/703/1/910
\bibitem[Pavlov et al.(1994)]{pav94} Pavlov, G.~G., Shibanov, Y.~A., Ventura, J., et al.\ 1994, \aap, 289, 837
\bibitem[Philippov et al.(2014)]{phi14} Philippov, A., Tchekhovskoy, A., \& Li, J.~G.\ 2014, \mnras, 441, 1879. doi:10.1093/mnras/stu591
\bibitem[Pires et al.(2019)]{pir19} Pires, A.~M., Schwope, A.~D., Haberl, F., et al.\ 2019, \aap, 623, A73. doi:10.1051/0004-6361/201834801
\bibitem[Posselt et al.(2013)]{pos13} Posselt, B., Pavlov, G.~G., Suleimanov, V., et al.\ 2013, \apj, 779, 186. doi:10.1088/0004-637X/779/2/186
\bibitem[Posselt \& Pavlov(2018)]{pos18} Posselt, B. \& Pavlov, G.~G.\ 2018, \apj, 864, 135. doi:10.3847/1538-4357/aad7fc
\bibitem[Posselt \& Pavlov(2022)]{pos22} Posselt, B. \& Pavlov, G.~G.\ 2022, \apj, 932, 83. doi:10.3847/1538-4357/ac6dca
\bibitem[Rajagopal \& Romani(1996)]{raj96} Rajagopal, M. \& Romani, R.~W.\ 1996, \apj, 461, 327. doi:10.1086/177059
\bibitem[Reynoso et al.(2006)]{rey06} Reynoso, E.~M., Dubner, G., Giacani, E., et al.\ 2006, \aap, 449, 243. doi:10.1051/0004-6361:20054236
\bibitem[Reynoso et al.(2017)]{rey17} Reynoso, E.~M., Cichowolski, S., \& Walsh, A.~J.\ 2017, \mnras, 464, 3029. doi:10.1093/mnras/stw2219
\bibitem[S{\'a}nchez-Ayaso et al.(2012)]{san12} S{\'a}nchez-Ayaso, E., Combi, J.~A., Albacete Colombo, J.~F., et al.\ 2012, \apss, 337, 573. doi:10.1007/s10509-011-0886-4
\bibitem[Spruit(2008)]{spr08} Spruit, H.~C.\ 2008, 40 Years of Pulsars: Millisecond Pulsars, Magnetars and More, 983, 391. doi:10.1063/1.2900262
\bibitem[Suleimanov et al.(2014)]{sul14} Suleimanov, V.~F., Klochkov, D., Pavlov, G.~G., et al.\ 2014, \apjs, 210, 13. doi:10.1088/0067-0049/210/1/13
\bibitem[Suleimanov et al.(2016)]{sul16} Suleimanov, V.~F., Poutanen, J., Klochkov, D., et al.\ 2016, European Physical Journal A, 52, 20. doi:10.1140/epja/i2016-16020-7
\bibitem[Suleimanov et al.(2017)]{sul17} Suleimanov, V.~F., Klochkov, D., Poutanen, J., et al.\ 2017, \aap, 600, A43. doi:10.1051/0004-6361/201630028
\bibitem[Suleimanov et al.(2012)]{sul12} Suleimanov, V.~F., Pavlov, G.~G., \& Werner, K.\ 2012, \apj, 751, 15. doi:10.1088/0004-637X/751/1/15
\bibitem[Suleimanov et al.(2010)]{sul10} Suleimanov, V.~F., Pavlov, G.~G., \& Werner, K.\ 2010, \apj, 714, 630. doi:10.1088/0004-637X/714/1/630
\bibitem[Thorstensen et al.(2001)]{tho01} Thorstensen, J.~R., Fesen, R.~A., \& van den Bergh, S.\ 2001, \aj, 122, 297. doi:10.1086/321138
\bibitem[Tian et al.(2008)]{tia08} Tian, W.~W., Leahy, D.~A., Haverkorn, M., et al.\ 2008, \apjl, 679, L85. doi:10.1086/589506
\bibitem[Tian et al.(2019)]{tia19} Tian, W.~W., Zhu, H., Zhang, M.~F., et al.\ 2019, \pasp, 131, 114301. doi:10.1088/1538-3873/ab35f4
\bibitem[Torres-Forn{\'e} et al.(2016)]{tor16} Torres-Forn{\'e}, A., Cerd{\'a}-Dur{\'a}n, P., Pons, J.~A., et al.\ 2016, \mnras, 456, 3813. doi:10.1093/mnras/stv2926
\bibitem[Vink et al.(1996)]{vin96} Vink, J., Kaastra, J.~S., \& Bleeker, J.~A.~M.\ 1996, \aap, 307, L41
\bibitem[Virtanen et al.(2020)]{vir20} Virtanen, P., Gommers, R., Oliphant, T.~E., et al.\ 2020, Nature Methods, 17, 261. doi:10.1038/s41592-019-0686-2
\bibitem[Vogt et al.(2018)]{vog18} Vogt, F.~P.~A., Bartlett, E.~S., Seitenzahl, I.~R., et al.\ 2018, Nature Astronomy, 2, 465. doi:10.1038/s41550-018-0433-0
\bibitem[Wijngaarden et al.(2019)]{wij19} Wijngaarden, M.~J.~P., Ho, W.~C.~G., Chang, P., et al.\ 2019, \mnras, 484, 974. doi:10.1093/mnras/stz042
\bibitem[Wijngaarden et al.(2020)]{wij20} Wijngaarden, M.~J.~P., Ho, W.~C.~G., Chang, P., et al.\ 2020, \mnras, 493, 4936. doi:10.1093/mnras/staa595
\bibitem[Willingale et al.(2003)]{wil03} Willingale, R., Bleeker, J.~A.~M., van der Heyden, K.~J., et al.\ 2003, \aap, 398, 1021. doi:10.1051/0004-6361:20021554
\bibitem[Wu et al.(2021)]{wu21} Wu, Q., Pires, A.~M., Schwope, A., et al.\ 2021, Research in Astronomy and Astrophysics, 21, 294. doi:10.1088/1674-4527/21/11/294
\bibitem[Zavlin et al.(1995)]{zav95} Zavlin, V.~E., Pavlov, G.~G., Shibanov, Y.~A., et al.\ 1995, \aap, 297, 441
\bibitem[Zavlin et al.(2000)]{zav00} Zavlin, V.~E., Pavlov, G.~G., Sanwal, D., et al.\ 2000, \apjl, 540, L25. doi:10.1086/312866




\end{thebibliography}

\end{document}